\documentclass[a4paper,fleqn,usenatbib]{mnras}

\usepackage{newtxtext,newtxmath}

\usepackage[T1]{fontenc}
\usepackage{ae,aecompl}

\usepackage{amsmath}	
\usepackage{amssymb}	
\usepackage{color}
\usepackage{epstopdf}
\usepackage[final]{graphicx}
\usepackage{float}
\usepackage[caption = false]{subfig}

\newcommand{\msun}{\mathrm M_\odot}
\newcommand{\hmsun}{h^{-1} \mathrm{M}_\odot}
\newcommand{\msunyr}{\mathrm M_\odot \rm ~yr^{-1}}
\newcommand{\kms} {\,{\rm km\,s}^{-1}}

\newcommand{\hmpc}{h^{-1}\,\mathrm{Mpc}}
\newcommand{\hkpc}{h^{-1}\,\mathrm{kpc}}
\newcommand{\hpc}{h^{-1}\,\mathrm{pc}}
\newcommand{\kpc}{\,{\rm kpc}}

\newcommand{\ac}{a_{\rm c}}
\newcommand{\aaa}{a_{30}}
\newcommand{\af}{a_{50}}
\newcommand{\Rh}{R_{\rm h}}
\newcommand{\Rspl}{R_{\rm spl}}

\title[ZOMG II: halo assembly \& galaxy formation]
{ZOMG II: Does the halo assembly history influence central galaxies and gas accretion?}

\author[E. Romano-D\'{\i}az et al.]{
Emilio Romano-D\'{\i}az\thanks{E-mail: emiliord@uni-bonn.de}, Enrico Garaldi\thanks{Member of the International Max Planck Research School (IMPRS) for Astronomy and Astrophysics at the Universities of Bonn and Cologne}, Mikolaj Borzyszkowski \& Cristiano Porciani
\\
Argelander Institut f{\"u}r Astronomie, Auf dem H{\"u}gel 71, D-53121 Bonn, Germany
}

\date{Accepted XXX. Received YYY; in original form ZZZ}

\pubyear{2016}

\begin{document}
\label{firstpage}
\pagerange{\pageref{firstpage}--\pageref{lastpage}}
\maketitle

\begin{abstract}
  The growth-rate and the internal dynamics of galaxy-sized
  dark-matter haloes depend on their location within the cosmic
  web. Haloes that sit at the nodes grow in mass till the present time
  and are dominated by radial orbits. Conversely, haloes embedded in
  prominent filaments do not change much in size and are dominated by
  tangential orbits.  Using zoom hydrodynamical simulations including
  star formation and feedback, we study how gas accretes onto these
  different classes of objects that, for simplicity, we dub
  `accreting' and `stalled' haloes. We find that all haloes get a
  fresh supply of newly accreted gas in their inner regions, although
  this slowly decreases with time, in particular for the stalled
  haloes. The inflow of new gas is always higher than (but comparable
  with) that of recycled material.  Overall, the cold-gas fraction
  increases (decreases) with time for the accreting (stalled)
  haloes. In all cases, a stellar disc and a bulge form at the centre of the
  simulated haloes. The total stellar mass is in excellent agreement
  with expectations based on the abundance-matching technique.  Many
  properties of the central galaxies do not seem to correlate with the
  large-scale environment in which the haloes reside.  However, there
  are two notable exceptions that characterise stalled haloes with
  respect to their accreting counterparts: i) the galaxy disc contains
  much older stellar populations; ii) its vertical scale-height is
  larger by a factor of two or more. This thickening is likely due to
  the heating of the long-lived discs by mergers and close flybys.
\end{abstract}


\section{Introduction}

Within the standard $\Lambda$CDM scenario, dark-matter (hereafter, DM)
haloes form in a hierarchical fashion. Matter falls onto them in two
distinct ways: via smooth accretion and by mergers. Baryons initially
follow the path dictated by the DM. However, they can radiate their
binding energy away and collapse into the central high-density regions
thus forming stars and subsequently galaxies \citep{white-rees78}.  In
this framework, the properties of galaxies are determined by the mass
and formation history of the DM haloes within which they form
\citep[e.g.][]{white-frenk91,somerville-primack99,cole+00}.
Semi-analytic models of galaxy formation are based on this concept
\citep[e.g.][]{bower+06,somerville+08,guo+11,benson12,henriques+15}
which is also naturally included in hydrodynamical numerical
simulations.

From the observational point of view, a galaxy-halo connection can
only be established at the statistical level. A commonly used approach
is the `abundance matching' technique, which is based on the
hypothesis that galaxy properties depend only on the mass of the host
halo
\citep[e.g.][]{kravtsov+04,vale-ostriker04,conroy+09,moster+13,behroozi+13a,behroozi+13,reddick+13}. This
approach is routinely used to infer the halo masses of galaxy
populations and inform models of galaxy formation
\citep{mandelbaum+06}, as well as to constrain cosmological parameters
from measurements of the large scale structure
\citep[e.g.,][]{tinker+05,cacciato+09,cacciato+13,more+13,villaescusa-navarro+14}

It has been shown that the clustering of haloes at fixed mass depends
on their formation history, the so-called halo assembly bias
\citep{sheth-tormen04,gao+05,li+08,dalal+08}. Internal properties
(e.g., halo concentration) also correlate with halo age
\citep{wechsler+06}.  

In order to explain the origin of this phenomenon, \citet[][heareafter
paper I]{miko+16} divided present-day galaxy-sized haloes into two
classes based on their mass-accretion history: those that are still
growing in size (accreting) and those that are not (stalled). Paper I
showed that accreting and stalled haloes populate different
environments of the cosmic web. Namely, stalled haloes reside within
prominent filaments (thicker than their sizes), while accreting haloes
at the nodes where thinner filaments converge. In addition, the
internal dynamics reflects the different accretion modes, i.e., the
velocity anisotropy profile is biased towards radial (tangential)
orbits in accreting (stalled) haloes.

If galaxy characteristics correlate with other halo properties than
the mass, then the existence of assembly bias would imply that the
abundance-matching method is inadequate. Observations
\citep{weinmann+06,kauffmann+13,knobel+15} and numerical simulations
\citep{chaves-montero+16,bray+16} provide evidence for such a
galaxy-halo correspondence.  For instance, neighbouring galaxies tend
to have similar colours, gas fractions and morphologies up to
separations of several Mpc, a phenomenon known as `galactic
conformity'.  These effects alter observables (i.e. two-point
correlations) at the 10-20 per cent level.  Therefore, it is
imperative to address them before the advent of the next-generation
wide-field galaxy surveys (e.g. eBoss, DESI, Euclid, and LSST) which
will produce extremely accurate measurements.  Attempts to extend the
abundance-matching technique have been proposed using either
age-matching \citep{hearin-watson13}, or a secondary halo property
beyond mass \citep{Hearin+16}.  In parallel with this phenomenological
approach, to make progress, it is necessary to shed new light on the
connection between galaxies and their host halos from the theoretical
point of view.

This article is the second in a series of three presenting a numerical
project named `Zooming On a Mob of Galaxies' (ZOMG).  The goal of the
ZOMG simulations is to investigate environmental effects on the
formation of present-day $L_*$ galaxies, their host haloes and
satellites.  In paper I we uncovered the physical origin of halo
assembly bias. In particular we used zoomed-in simulations of stalled
and accreting haloes to study their assembly history and structural
properties. In this paper we consider a sub-set of the Dark-ZOMG
(DM-only) runs and resimulate them adding the baryonic
component, including star formation, cooling and feedback. Our aim is
to characterise gas accretion and the properties of the central galaxy
in haloes with different formation histories.  In paper III
\citep{garaldi+16} we address the connection with the satellite galaxy
population.

From the theoretical point of view, gas inflow in haloes with masses
of $10^{11.6} \msun$ has been shown to be bimodal, the so-called cold
and hot accretion
~\citep{birnboim-dekel03,keres+05,dekel-birnboim06,keres+09a}. Haloes
below this threshold are not massive enough to support shocks.  In
this case, cold gas reaches the halo centre along filaments, feeding
directly the main galaxy. On the other hand, in more massive hosts the
gas is shock heated to (and above) the virial temperature of the
halo. The gas smoothly fills the halo until it cools down and rains
into the galaxy, thus sustaining its star-formation rate. Cold
accretion dominates at high redshifts ($z>2$) while the hot mode
becomes more efficient at later epochs \citep{keres+05}.

A natural question that raises at this point is: how is the gas
accretion scenario coupled with the stagnation or growth of the
underlying DM haloes?  Moreover, how does the halo assembly history
affect the formation and evolution of the central galaxies?  Given
that stalled haloes do not grow in mass since long ($\sim$10 Gyr),
should one expect them to host red and passive galaxies by the present
time? More generally, is there a galaxy property that correlates with
halo formation time and generates galaxy assembly bias?  In this paper
we address these issues. For this purpose we select and re-simulate
two acccreting and two stalled haloes from the Dark-ZOMG suite. We
study gas accretion at different physical radii ranging from distances
well beyond the halo radius all the way down to the central
galaxy. Results are contrasted with the inflow rate of dark matter at
the same locations. We investigate the morphological structure of the
central galaxies in terms of their stellar density profile (both
radial and vertical) and kinematic decomposition.

The outline for this paper is as follows. In
Section~\ref{sec:simulations} we describe our numerical simulation
setup, the identification of haloes and galaxies, define their
formation times, and the construction of their merger trees.  In
Section~\ref{sec:results} we present our main results.  In
Section~\ref{sec:Raccretion} we study the mass accretion rates for
both DM and gas. The analysis of the central galaxies is presented in
Section~\ref{sec:gals}, their comparison with observations in
Section~\ref{sec:observations} and their morphological
characterisation in Section~\ref{sec:decomposition}. A non-parametric
analysis of their stellar distributions is shown in
Section~\ref{sec:cas}. Finally, we present our conclusions in
Section~\ref{sec:conclusions}.

\section{Numerical Simulations and methodology}
\label{sec:simulations}

We have made use of the ZOMG simulations presented in paper I. This
simulation-set follows the evolution of seven zoomed-in DM haloes that
have been chosen according to their collapse time. In the present
paper we are interested in studying how the halo gas content is
affected by the difference in their formation time, and in turn, how
this affects the evolution of the central galaxies. For these
purposes, we have extended part of such simulation suite as to include
baryons in four of those haloes. We have chosen two that have
collapsed before $z \sim 1$ and therefore, their respective matter
accretion has been `stalled'. The other two chosen haloes are still
`accreting' matter at the present epoch and have not fully collapsed
yet. Following the  nomenclature used in paper I, the selected haloes are:
Supay, Siris, Amun and Abu, where the initial letter stands for
stalled (S) or accreting (A). The S-haloes have the particularity that
they are immersed within filaments, while the A-haloes are residing at
the knots (intersections) of two or more filaments (see Fig. 9 in
paper I).

For a full explanation about the simulation setup and selection of the
haloes we refer the reader to paper I. In this section we specify
their main characteristics and the inclusion of the baryonic
component.

\subsection{The code}

We have used a modified version of the tree-particle-mesh smoothed
particle hydrodynamics (SPH) code {\sevensize{PGADGET-3}}
\citep{gadget}, using the conservative entropy formulation
\citep{springel-hernquist+02}. Our implementation includes radiative
cooling by H, He, and metals, star formation, stellar feedback,
constant-velocity galactic winds, a uniform ultraviolet (UV) background
\citep{haardt-madau01} and a sub-resolution model for the multiphase
interstellar medium, where SPH particles contain both
cold and hot phases \citep{yepes+97,springel-hernquist+03}.  Since our haloes
have masses $M_{\rm h}\sim 5\times 10^{11} \msun$, we do not include
AGN feedback which is thought to regulate star formation in more
massive objects \citep[see][and references
therein]{dimatteo+05,sijacki+07,fabian12,ishibashi+12,cresci+15}.
Star formation can only take place in regions where the gas density
exceeds the threshold $n_{\rm H,SF}=0.1 \,{\rm cm^{-3}}$. This is
calibrated to reproduce the observed Kennicutt-Schmidt relation
\citep{kennicutt98} at $z=0$. Each gas particle has an assigned
star-formation rate (SFR). However, the actual conversion from gas to
stars proceeds stochastically, where each star particle takes half of
the initial gas mass \citep{springel-hernquist+03}.

\subsection{Setup}

The D-ZOMG set presented in paper I follows the evolution of DM haloes
with masses around the peak in the expected stellar mass - halo mass
relation at the present time \citep{behroozi+13,moster+13}.

The suite was drawn from a low resolution, collisionless
(DM-only) parent simulation. The initial conditions (ICs) were
generated with the {\sevensize{MUSIC}} code \citep{music}. The ICs
were imposed on a $50 \, \hmpc$ box size, with a resolution of $512$
grid cells per dimension. The cosmology adopted was that of the
\cite{planck+14}, i.e. $\Omega_{\rm m} = 0.308$,
$\Omega_{\Lambda}=0.692$, $\Omega_{\rm b} = 0.0481$, $h=0.678$.  The
linear power spectrum was characterised by the spectral index
$n=0.9608$ and normalised by the linear rms amplitude of mass
fluctuations in spheres of $8 \, \hmpc$ radius, $\sigma_8 =
0.826$. The simulation was evolved from $z=99$ up to $z=0$.

The collapse time for all haloes was calculated following
\cite{miko+14} (see Section~\ref{sec:assembly}). Seven haloes with mass
$M_{\rm h} \sim 5\times 10^{11} \msun$ and different collapse times
were selected for re-simulation at higher resolution. In order to
avoid contamination from massive particles in the multi-mass
re-simulated haloes, all particles within three virial radii at $z=0$
were selected.  These particle distributions were traced back to the
ICs.  The corresponding comoving
volumes were resampled at higher resolution with
{\sevensize{MUSIC}}. The global setup includes six nested levels of
refinement and periodic boundary conditions. To minimise the
computational effort in the zoom-in simulations, a minimal ellipsoid
was carved out containing all the particles that by $z=0$ would end up
within the three virial radii.

For the present study, we have modified the ICs of our chosen four
haloes as to add baryons only at the highest level of refinement. The
final effective resolution in the high-resolution region was $4096^3$
in both DM and SPH particles, with a gravitational softening of
$\epsilon = 240 \hpc$ (comoving). The corresponding particles masses
are $1.31 \times 10^5\hmsun$ in DM and $2.43 \times 10^4 \hmsun$ for
gas. Since gas particles could experience up to two episodes of star
formation (SF), the resulting stellar masses are $1.21\times 10^4
\hmsun$.

We imposed a fine time interval of 20 Myr between outputs to perform
accurate post-processing analysis of the evolution of the
structures. There are in total 682 snapshots for each halo.

\subsection{Halo and galaxy identification}
\label{sec:haloes}

Haloes were defined as gravitationally bound objects by means of the
Amiga Halo Finder code {\sevensize{AHF}} \citep{Gill+04,ahf}. They
were initially identified as those spherical regions with mean density
equal to $200 \rho_{\rm c}(z)$, with $\rho_{\rm c}(z)$ being the
critical density of the universe at a given redshift.  After this, an
iterative unbinding procedure was applied by removing particles with
velocities larger than $1.5 \, \varv_{\rm esc}$, where $\varv_{\rm
  esc}$ is the escape velocity of the halo.  In the case of gas
particles, their thermal energy was also taken into
consideration. Halo masses and radii ($M_{\rm h}$ and $\Rh$,
respectively) were computed from the bound distributions. In
particular, $\Rh$ was defined as the minimum spherical radius
containing all the bound particles. Substructures and possible
sub-substructures were identified along the same lines (see paper
III). Halo quantities such as spin, shape, number of substructures,
among others, were computed within these distributions. Note that we
followed the same formalism to identify haloes as in paper I, with the
only difference that our haloes are composed by DM and baryons (gas
and stars).  The main physical properties of our halo sample are shown
in Table~\ref{tab:haloes}. Differences in these quantities from paper
I are due to the inclusion of baryons.

It is worth noticing that the definition of a halo depends on the
critical matter density of the universe.  Such definition can lead to
a spurious pseudo-evolution of halo mass and size because of the
redshift-dependence of the reference density, even if its physical
density profile remains constant over time
\citep{diemer+13,diemer+14,zemp14,adhikari+14}. As shown in paper I,
it is difficult to draw a distinctive boundary between a halo and its
environment since the transition is smooth. \cite{adhikari+14} and
\cite{more+15} suggested the `splashback' radius as that region which
is not subject to pseudo-evolution.  They associated the steepening in
the radial density profile of haloes with the apocentre of accreted
matter which is on its first orbit.  We have included the splashback
radius ($\Rspl$, hereafter) values at $z=0$ in Table~\ref{tab:haloes}
(see paper I for details).

The main galaxy for each halo was defined as the central region
containing both gas and stars within a radius $r_{\rm gal}=0.1\,\Rh$
\citep[e.g.][]{Scannapieco+12}. 

\begin{figure}
  \includegraphics[width=\columnwidth]{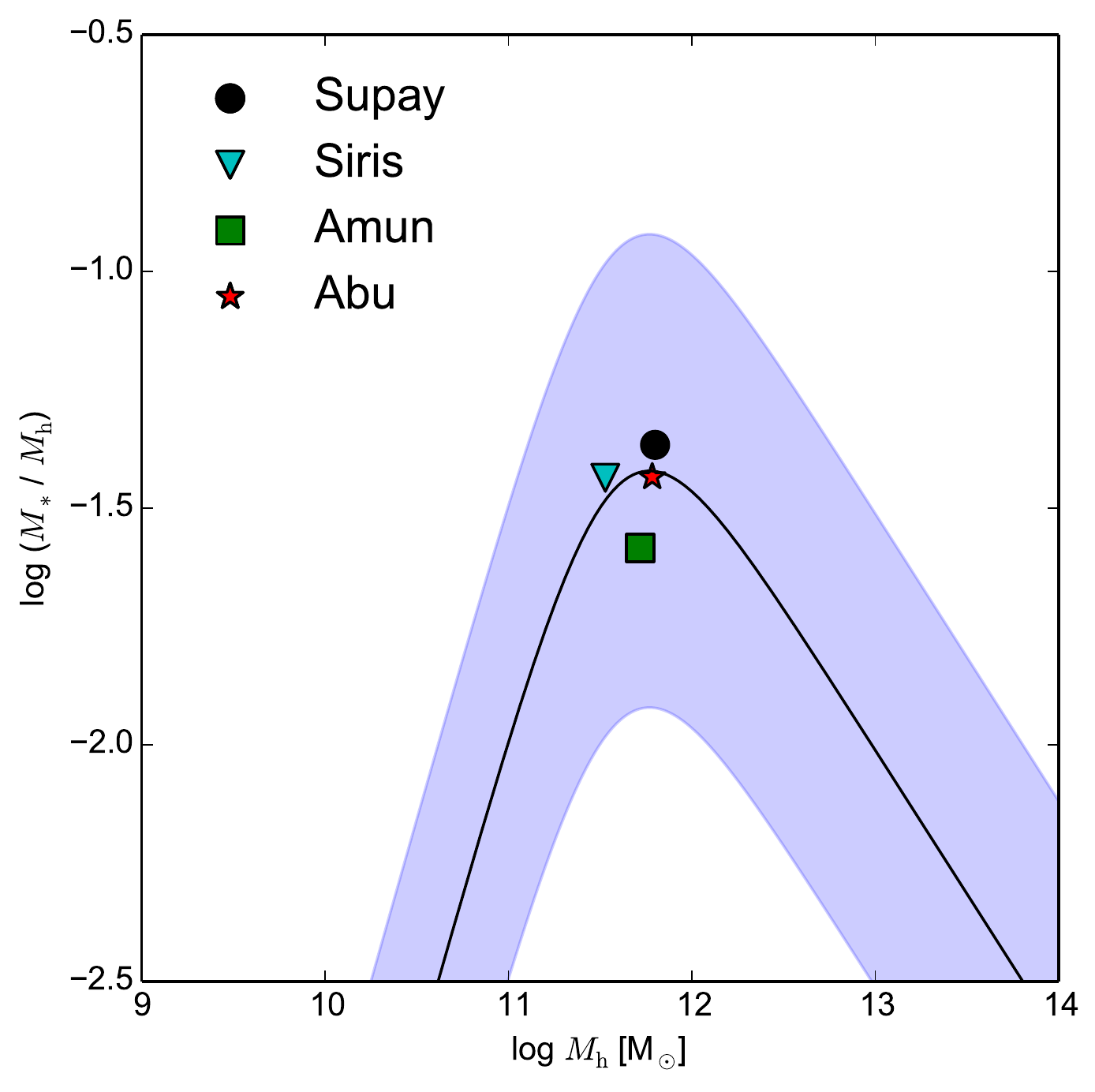}
  \caption{Stellar mass -- halo mass relation at $z=0$. The different symbols
    indicate our four haloes.  The black line
    represents the expected trend from \protect\cite{moster+13} while
    the blue-shadowed region the $\pm 1\sigma$ uncertainty around the
    mean value.}
  \label{fig:smhm}
\end{figure}

We have checked that our haloes and galaxies are in agreement with the
expected $M_{\rm h} - M_*$ relation \citep{behroozi+13,moster+13} at
$z=0$. Fig.~\ref{fig:smhm} shows the stellar-mass fractions as a
function of halo mass for our simulation set.  Notice that all of our
galaxies agree very well with the expected trend (black solid line)
and with the peak of the distribution.

\begin{figure}
  \includegraphics[width=\columnwidth]{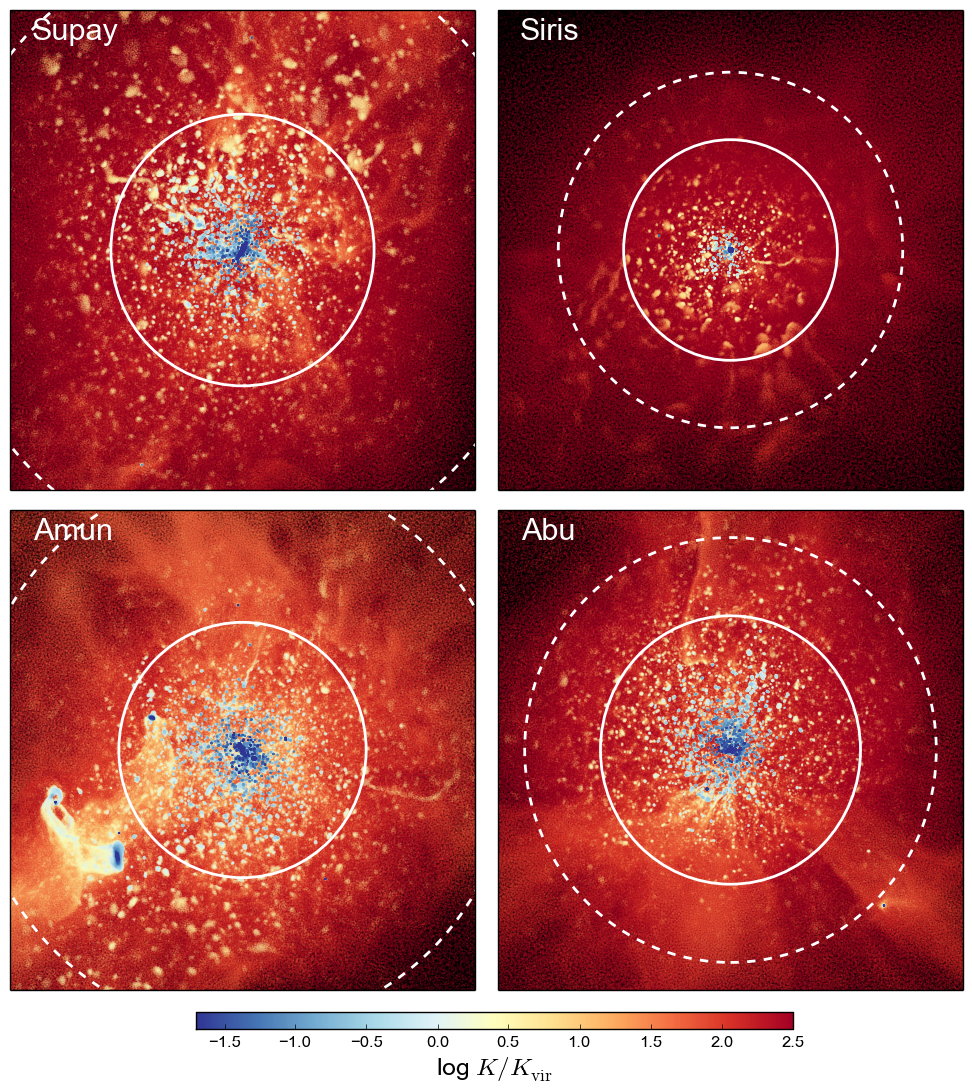}
  \caption{Gas distribution around our re-simulated haloes at
    $z=0$. Gas particles have been colour-coded according to their
    entropy normalised by the virial entropy of their respective
    haloes. The panels refer to cubes of length of 540 $h^{-1}$kpc
    projected along one of the axes The solid circles indicate $\Rh$,
    while the dashed ones $\Rspl$. }
\label{fig:halos-gas}
\end{figure}

The gas distributions in and around our haloes are illustrated in
Fig.~\ref{fig:halos-gas}, where 2D projections are shown at
$z=0$. Each panel refers to a cube with side length of 540 $h^{-1}$kpc
(same as in Fig. 4 of paper I) projected along one of the axes. The
solid circles represent the halo radius ($\Rh$), while the dashed ones
the splashback radius ($\Rspl$).  Particles have been colour-coded
using their local `entropy' \citep[$K \propto T/\rho^{2/3}$,
see][]{dekel+09} in units of the virial temperature and mean density
within $\Rh$.  The figure highlights the structure and kinematics of
the gas distributions. The entropy maps exhibit gas heated by virial
shocks indicated by red tones that cover all the surroundings of the
stalled haloes, and to a lesser degree the accreting ones. In the case
of Siris, this extends even beyond $\Rspl$. The A-haloes show some
streams of lower entropy (light colours) penetrating into the central
parts. They represent cold gas being accreted into the central regions
(see Section~\ref{sec:gas}). The boundaries between streams and the
hot intergalactic medium seem to be well defined. The main galaxies
(and some substructures) are depicted as regions of very low entropy
indicated by the blue tones.

\subsection{Formation and collapse times}
\label{sec:assembly}

\begin{figure}
  \includegraphics[width=\columnwidth]{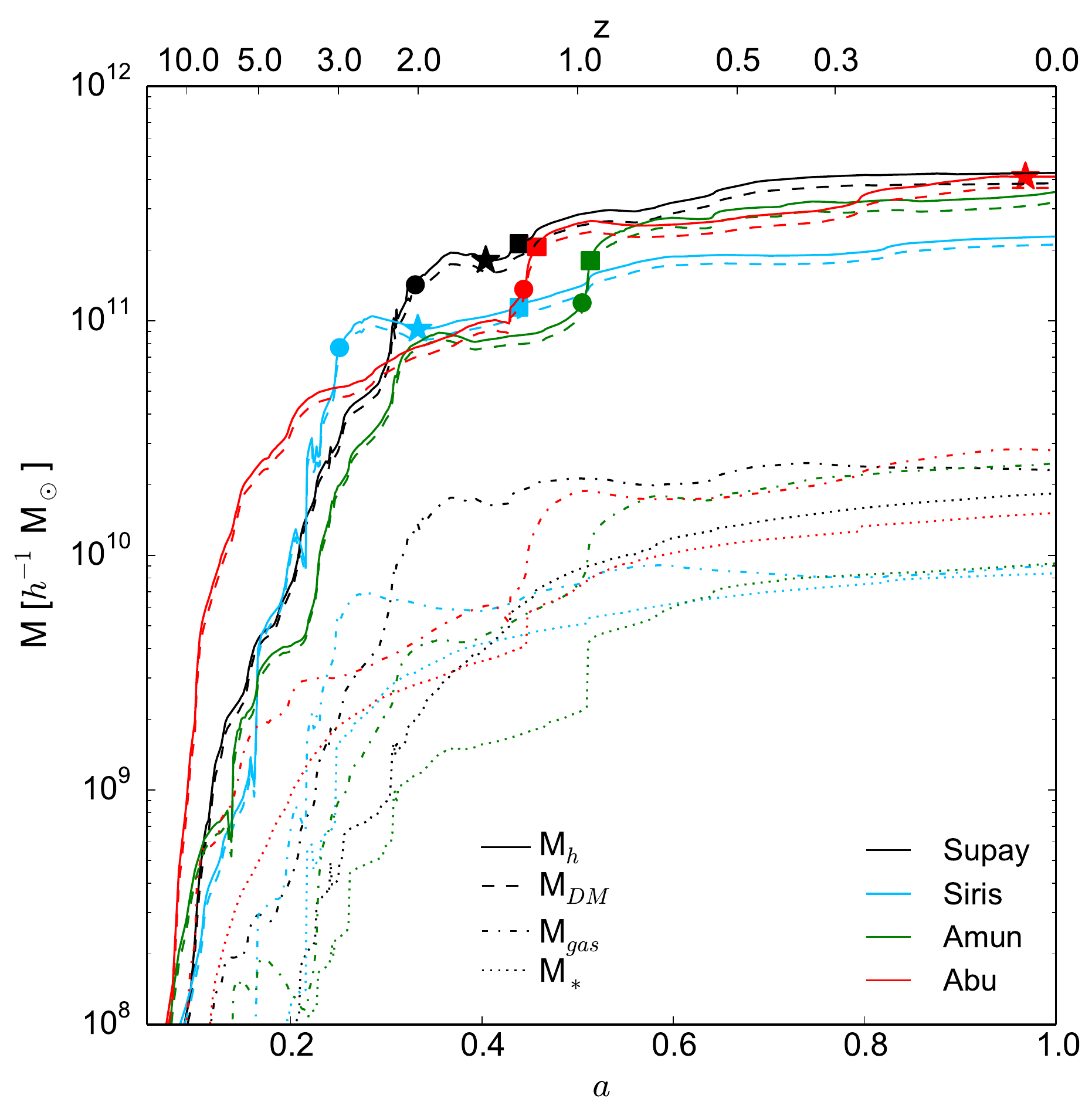}
  \caption{Mass-accretion histories for our haloes. The different
    lines represent the contributions from the three mass components:
    DM (dashed line), gas (dot-dashed) and stellar (dotted lines),
    while the continuos lines indicate the total matter
    evolutions. The symbols indicate the times at which the haloes
    have accumulated 30 per cent ($\aaa$, circles) and half ($\af$,
    squares) of their final masses, while the stars indicate the
    haloes collapse times $\ac$.}
    \label{fig:mahs}
\end{figure}

We computed the mass accretion histories (MAHs) of our haloes,
following the main branch of their merger tree.
A parent halo was defined as the halo between two neighbouring time
outputs $(t_i,t_j)$ that maximises the merit function $(N^2_{i \cap j}
/N_iN_j)$, where $N_i$ and $N_j$ are the numbers of particles within
the halo at time $t_i$ and its progenitor at time $t_j$, respectively,
while $N_{i \cap j}$ is the number of particles shared between both
times \citep{ahf}.
Merger trees were traced from $z=0$ until the time their corresponding
parent haloes could be identified by
{\sevensize{AHF}}. Fig.~\ref{fig:mahs} shows the MAH for our
haloes: continuous, dashed, dot-dashed and dotted curves represent the
total, DM, gas and stellar masses, respectively.  Haloes are
identified with different colours. 

We defined halo formation time in different ways \citep[see
also][]{li+08}. 
We denoted by $\af$ ($\aaa$) the earliest time at which half (30 per
cent) of the mass is in place
(filled squares and circles in Fig.~\ref{fig:mahs}). 
We also computed the collapse time $\ac$ (stars in
Fig.~\ref{fig:mahs}) as in \citet[][see also paper I]{miko+14}. This
quantity identifies the epoch at which the physical volume filled by
the forming haloes becomes stable and virialization can happen. For
Supay and Siris this happened at $\ac < 0.5$, while for Abu and Amun
$\ac \gtrsim 1$. [We recall that the definition of stalled and
accreting haloes is based on $\ac$]. Notice that for the stalled
haloes $\ac$ and $\af$ are very similar, with the former taking place
before the latter.  The situation for the accreting-haloes is
substantially different, as they reach their $\af$ by $z \sim 1$ and
$\ac$ by $z \lesssim 0$.

Overall, these times are very similar to those reported in paper I,
the small differences are due to the effect of baryons that slightly
affect the collapse time of the main structures as well as the
incoming ones.

\begin{table*}
  \centering
  \caption{Halo properties at $z=0$. The first column denotes the halo name. 
    Names starting with `S' refer to stalled haloes while those starting 
    with `A' to accreting haloes. The second column represents the mass of 
    the halo, while the third column their radius. The fourth column shows the 
    splashback radius. The halo spin parameter \citep{bullock01ck} is shown 
    in the fifth column. The sixth and seventh columns represent the times (in
    terms of the expansion factor $a$) at which the haloes first reached a 
    third ($\aaa$), and a half ($\af$) of their final mass. The eighth column 
    shows the collapse time ($\ac$) defined as in \protect\cite{miko+14}. 
    The ninth column indicates the number of resolved substructures within each halo.}
  \label{tab:haloes}
  \begin{tabular}{lcccccccc}
    \hline
    Halo name & $M_{\rm h}$ & $\Rh$  & $\Rspl$ & $\lambda$ & $\aaa$ & $\af$ & $\ac$ & \# substructures \\
                         & [$10^{11}\hmsun$] & [$\hkpc$] & [$\hkpc$] & & & & & \\
    \hline
    Supay & 4.3 & 153 & 348 & 0.0284 & 0.330 & 0.439 & 0.404 & 630 \\
    Siris & 2.3 & 124 & 200 & 0.0087 & 0.251 & 0.439 & 0.333 & 228 \\
    Abu   & 4.1 & 151 & 239 & 0.0163 & 0.444 & 0.458 & 0.968 & 364 \\
    Amun  & 3.5 & 144 & 314 & 0.0347 & 0.505 & 0.513 & >1.000 & 457 \\
    \hline
  \end{tabular}
\end{table*}

\subsection{Accretion rates}
\label{sec:accretion}

Once the main branches of the merger trees were identified, we
calculated the matter inflows and outflows as a function of time at
given radii. We have chosen to work with physical rather than comoving
coordinates.  In order to calculate the accretion, we have tagged all
particles (DM and baryons) that are or were part of the halo at a
given time. We registered the time at which each particle became part
of the halo or of a region of interest within the halo. With this
list, and by comparing the matter content within the given region
between two neighbouring output times, one can distinguish if a
particle has been accreted for the first time, re-accreted or if it
has been expelled \citep[see also][]{erd+14}.  The fine output-time
sampling of our simulation suite allows us to have a very detailed and
precise measurement of such inflows and outflows.
Notice that our approach differs from other techniques where accretion
is defined by differentiating the MAH 
\citep[e.g.][]{murali+02,maller+06,keres+09a,fakhouri+10,vandevoort+11},
or by measuring the flux of matter at fixed radii
\citep[e.g.][]{fauchergiguere+11}.

\section{Results}
\label{sec:results}

\begin{figure}
  \includegraphics[width=\columnwidth]{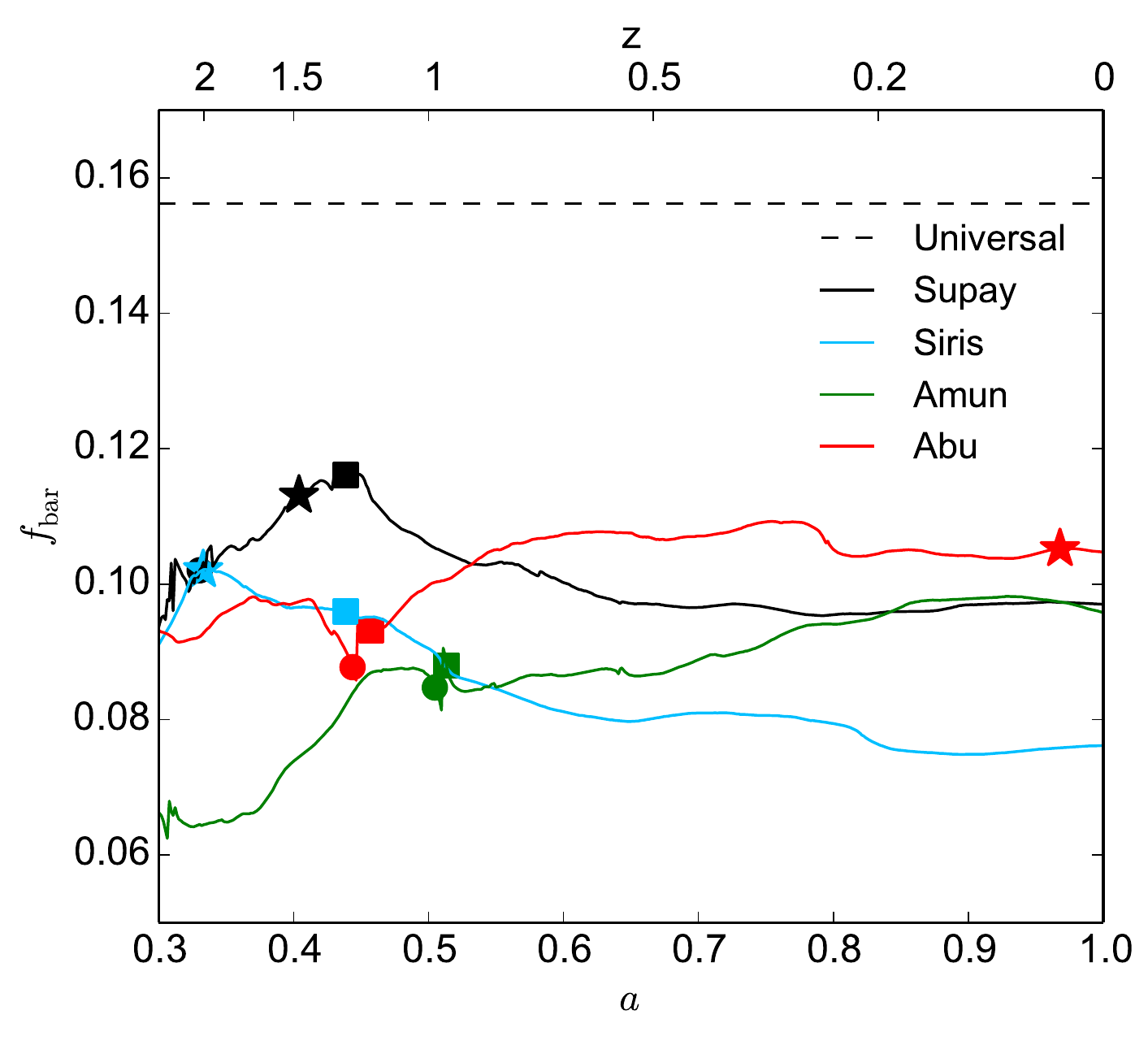}
  \caption{Halo baryon fraction evolution for our simulation
    suite. Colours indicate different haloes. The dashed line
    represents the universal baryon fraction. Symbols are the same as
    in Fig.~\ref{fig:mahs}, i.e., circles represent $\aaa$, squares
    $\af$ and stars $\ac$.}
    \label{fig:fbar}
\end{figure}

In Fig.~\ref{fig:fbar} we analysed how the baryonic mass fraction
($f_{\rm bar} = M_{\rm gas + stars} / M_{\rm h}$) within $R_{\rm h}$
changes with time.  To establish a link with the MAHs, we indicated
the different characteristic times of the haloes using the same
convention as in Fig.~\ref{fig:mahs}.  As a consequence of feedback
processes, $f_{\rm bar}$ is always lower than the universal baryon
fraction (dashed line) and shows relatively little variations.  This
is consistent with previous numerical results
\citep[e.g.][]{crain+07,fauchergiguere+11,Scannapieco+12,peirani+12,brook+14,metuki+15}
and observations \citep{mcgaugh+10}.  The nearly constant value of
$f_{\rm bar}$ over time implies that the baryon mass grows
approximately at the same rate as the DM.

On top of these general trends, a clear distinction between early- and
late-collapsing haloes is noticeable: the stalled ones show a slight
decrease in $f_{\rm bar}$ after having reached their maximum baryonic
fraction (in the case of Supay this time coincides with $\af$, while
for Siris with $\aaa$, both of these being very close to $\ac$).  At
later times, $f_{\rm bar}$ stays almost constant.  On the other hand,
the baryon fraction of accreting haloes continues raising slowly until
$a \approx 0.8$.

\begin{figure}
  \includegraphics[width=\columnwidth]{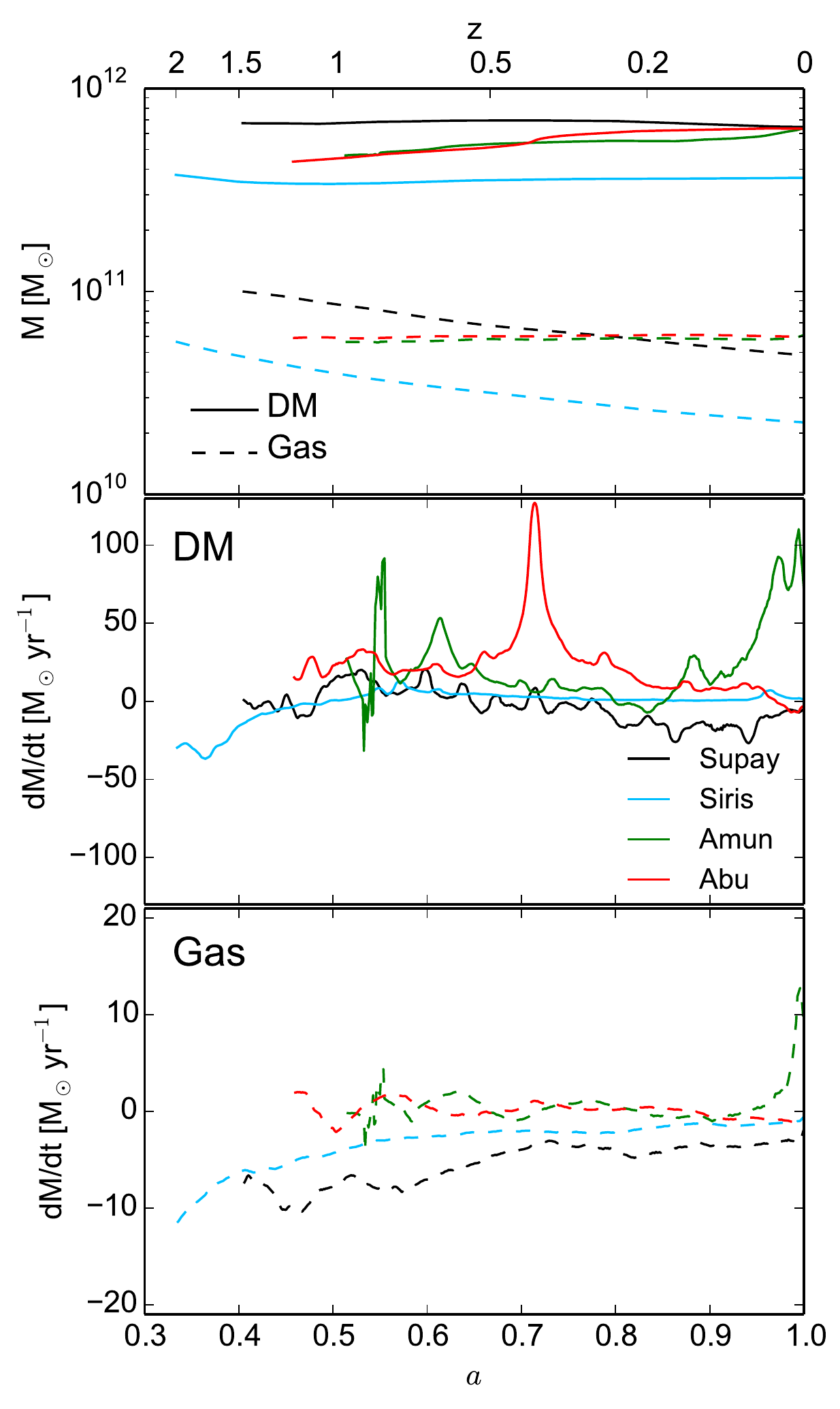}
  \caption{DM and gas content evolution within spheres of $300 \kpc$
    from the halo centres (top panel). This radius is of the order of
    the halo splashback radius. The middle and lower panels indicate
    the net mass accretion rate at the same radius.}
    \label{fig:rsplash}
\end{figure}

\begin{figure}
  \includegraphics[width=\columnwidth]{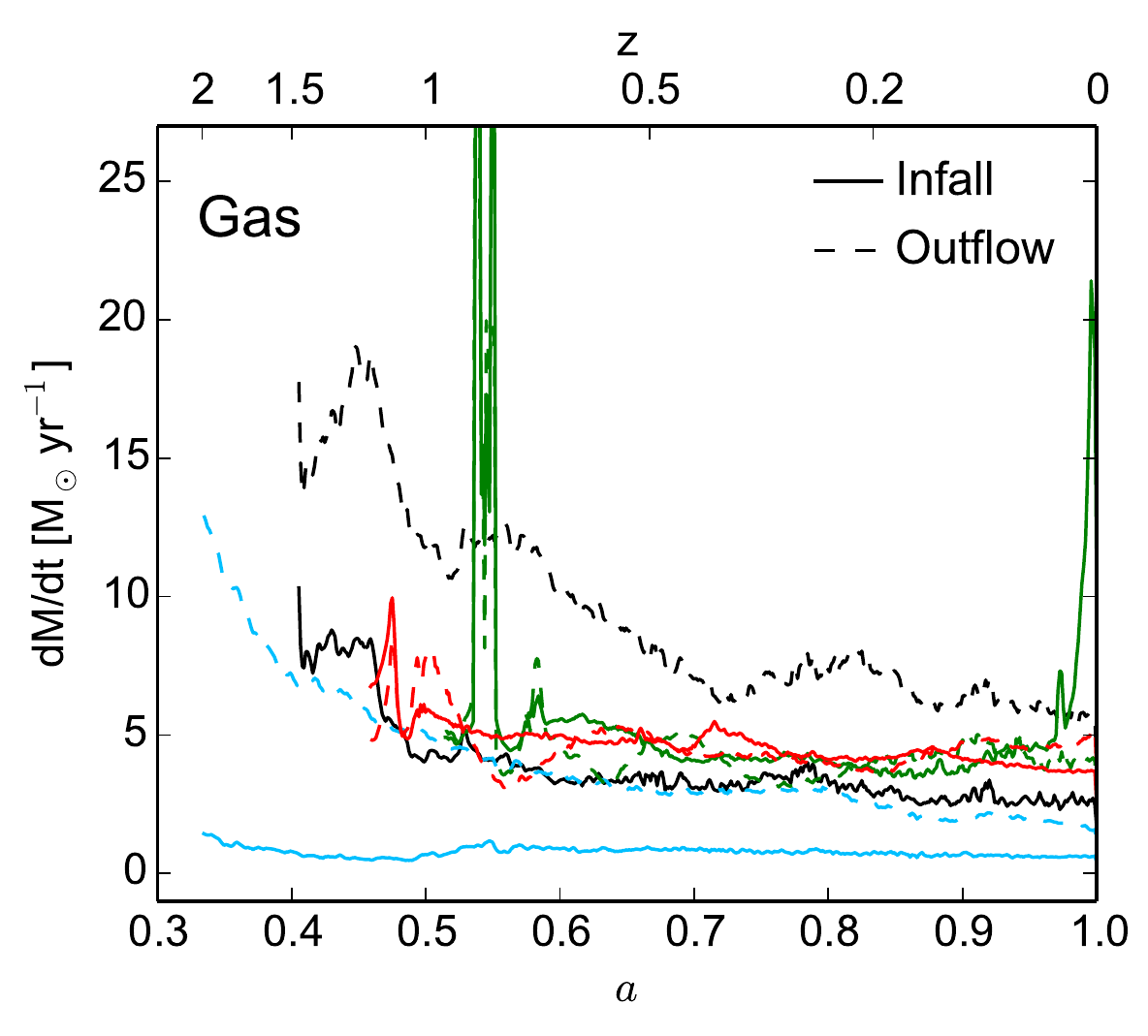}
  \caption{Gas infall (solid lines) and outflow (dashed lines) rates
    for our halo suite at $r_{300}$.}
    \label{fig:rsplash_in-out}
\end{figure}

\subsection{Accretion}
\label{sec:Raccretion}

In this section we describe how matter was accreted onto the haloes as
a function of time. The starting point for our calculations was set as
$a_{\rm i} = \min( \af,\ac)$ so that to analyse a considerable
interval for each halo.

\subsubsection{Halo outskirts}

We first explored the outermost regions of the haloes.  In order not
to suffer from the pseudo-evolution, we calculated the matter inflow
and outflow rates at fixed physical radii rather than using
conventional choices such as $\Rh$ or $r_{\rm gal}$.  In
Fig.~\ref{fig:rsplash} we show the evolution of the matter content
calculated at fixed radii of 300~kpc ($r_{300}$), a distance
comparable to $\Rspl$.  The top panel shows the DM and gas masses
while the central and lower panels display the corresponding time
derivatives. In Fig.~\ref{fig:rsplash_in-out} the gas infall and
outflow rates are presented.
For the S-haloes, the DM content does not evolve after $\ac$, while
the gas mass nearly halves by the present time.  This is mostly due to
the combined action of gas consumption via star formation and gas
ejection via feedback processes which are more efficient than the
ability of the haloes to capture gas from their surroundings.  In
fact, outflows are considerably larger (by almost a factor of two)
than the corresponding inflows (see bottom panel in
Fig.~\ref{fig:rsplash} and Fig.~\ref{fig:rsplash_in-out}).
On the other hand, the accreting haloes increase their DM mass by a
factor of two, while their gas content remains nearly constant given
that infall and outflow rates cancel each other. Notice that the
infall rates are larger than in the S-haloes.

\subsubsection{Dark-matter accretion within haloes}
\label{ss:dm}

In order to study the accretion onto the halo and how gas flows in its
inner region, we considered an outer radius of $100\kpc$
($r_{100}$)\footnote{Choosing a larger radius, such as $150\kpc$ or
  more does not change qualitatively our results.} and an inner radius
of $20\kpc$ ($r_{20}$, comparable to the size of the central
galaxies).

\begin{figure}
  \includegraphics[width=\columnwidth]{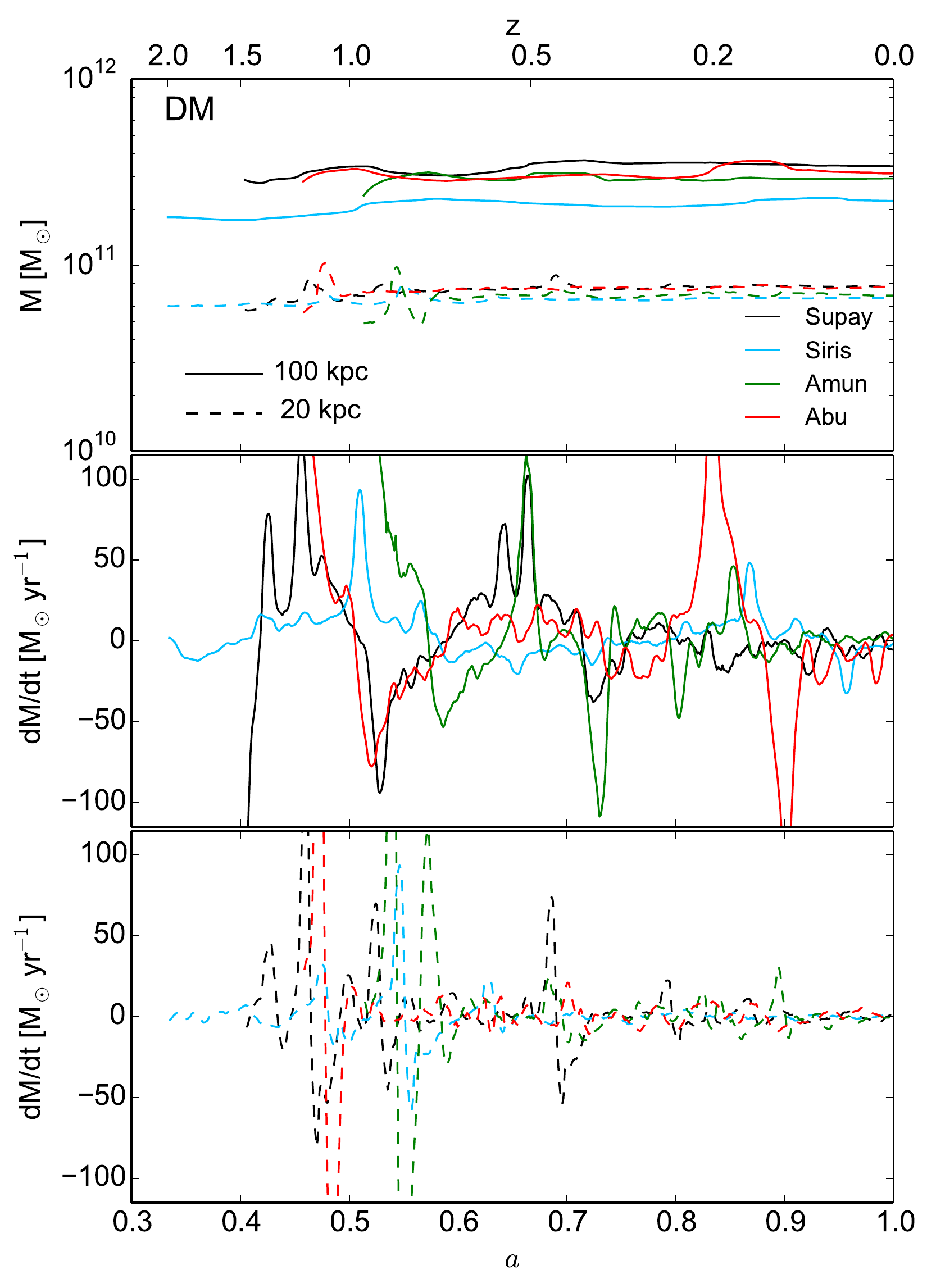}
  \caption{Top-panel: DM content evolution for our halo-suite
    calculated at $r_{100}$ (solid lines) and $r_{20}$ (dashed
    lines). Different colours indicate different simulations.  The central
    and bottom panels show the net mass accretion at the same radii.}
  \label{fig:dm_r}
\end{figure}

\begin{figure*}
  \includegraphics[width=7in]{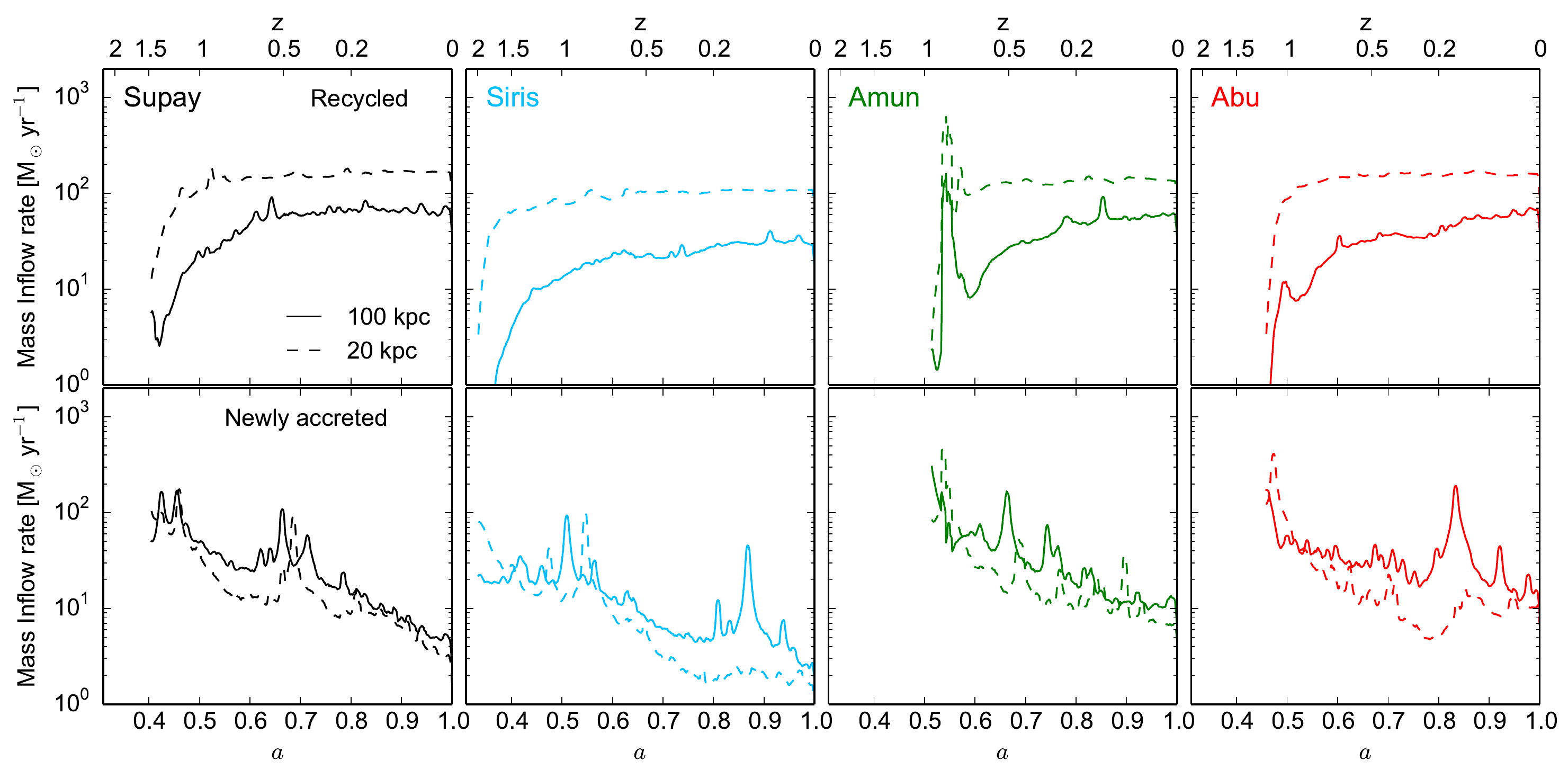}
  \caption{Inflow DM accretion rates for our simulated halo set at two
    different radii, $r_{100}$ (continuos lines) and $r_{20}$ (dashed
    lines). The top panel depicts the recycled DM accretion while the
    bottom panel indicates the newly accreted material (material being
    accreted at a given radius for the first time).}
  \label{fig:dm_accretion}
\end{figure*}

Fig.~\ref{fig:dm_r} shows the evolution of the DM content (top panel)
within $r_{100}$ (solid lines) and $r_{20}$ (dashed lines) for
all our haloes. All masses grow only by $10-20$ per cent, showing that
in these regions inflows and outflows are approximately in
equillibrium (see bottom panel of Fig.~\ref{fig:dm_r}).  The temporary
deviations, denoted by local oscillations in the masses indicate
interactions with substructures. However, these events do not add any
significant amount of material.

In Fig.~\ref{fig:dm_accretion} we have analysed in more detail the
infalling material dividing it into recycled (upper panels) and newly
accreted matter (bottom panels).  Recycled material refers to those
particles that have been previously accreted onto the selected volumes and
due to various processes (i.e. radial migration, encounters, mergers)
have been expelled and have fallen back into them.  For all haloes,
the rate at which material is recycled at $r_{20}$ is higher by a
factor of two than at $r_{100}$.  This indicates that there is DM
which is confined within the inner regions of the halo and has limited
radial excursions \citep[e.g.,][]{erd-I}. Notice that the amount of
recycled material at $r_{100}$ increases with time for all haloes but
it saturates earlier for the stalled ones ($z\sim0.6$), while it
continues increasing until the present time for their accreting
counterparts.

Typically, only a fraction ($\sim$50 per cent or less) of the newly
accreted material that infalls into $r_{100}$ reaches $r_{20}$.
Overall, the infall rates for the S-haloes are lower than those in the
A-haloes at both radii. This becomes more evident at lower redshifts
where the differences go up to a factor of 5 between the two types of
haloes, and it is independent of the chosen radius.  Spikes (local
maxima) along the evolution are signatures of incoming substructures
orbiting around and within the halo\footnote{These can also be seen in
  the recycled material although to a lesser degree due to
  substructure destruction (see also paper III).}.  Some of these
cross both $r_{100}$ and $r_{20}$ during their first passage with a
time lag.  However, most of them, do not pass close enough to the
centre as shown by the relative lack of peaks in the $r_{20}$ dashed
lines

\begin{figure}
  \includegraphics[width=\columnwidth]{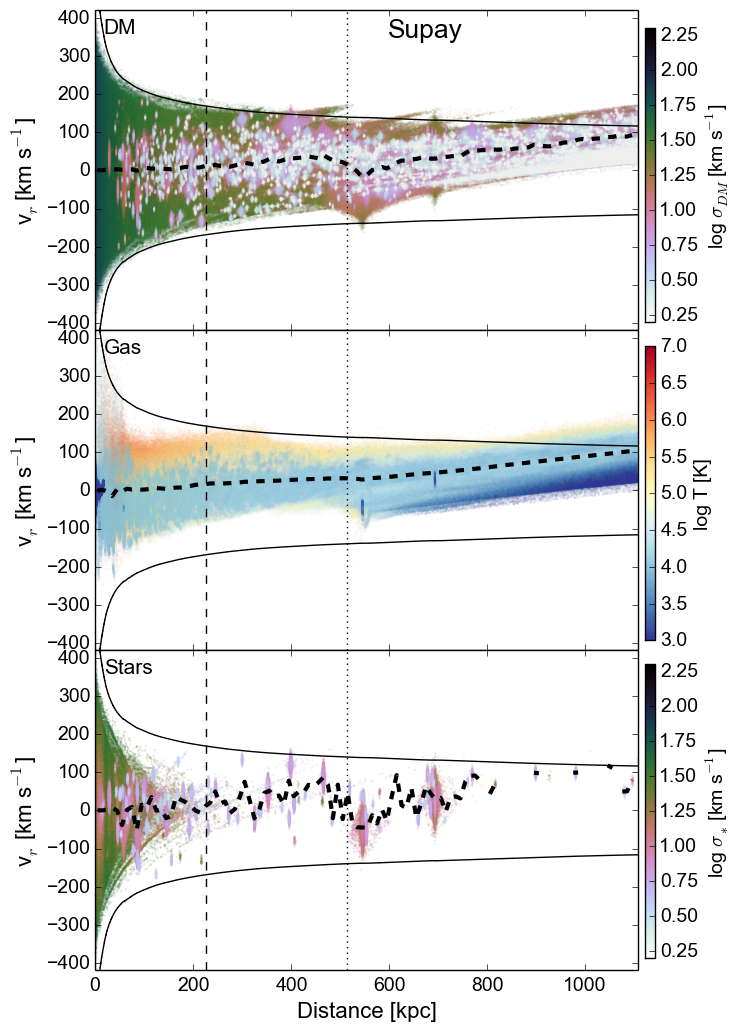}
  \caption{Radial phase-space diagram for Supay at $z=0$. Each panel
    shows a different component: DM (top), gas (middle) and stars
    (bottom). The colour coding is indicated by the right bars and is
    directly proportional to the local velocity dispersions
    $\sigma_{\varv}$ (DM and stars) and temperature $T$ (gas). The
    dashed-vertical lines indicate the halo radius $\Rh$, the
    dotted-vertical lines the splashback radius $\Rspl$. The thick
    dashed curves represent the average mass-weighted radial velocity
    $\varv_r$ as a function of distance, while the solid black lines
    above and below the distributions show the corresponding escape
    velocity as a function of distance.}
  \label{fig:phase_space-s}
\end{figure}

\begin{figure}
  \includegraphics[width=\columnwidth]{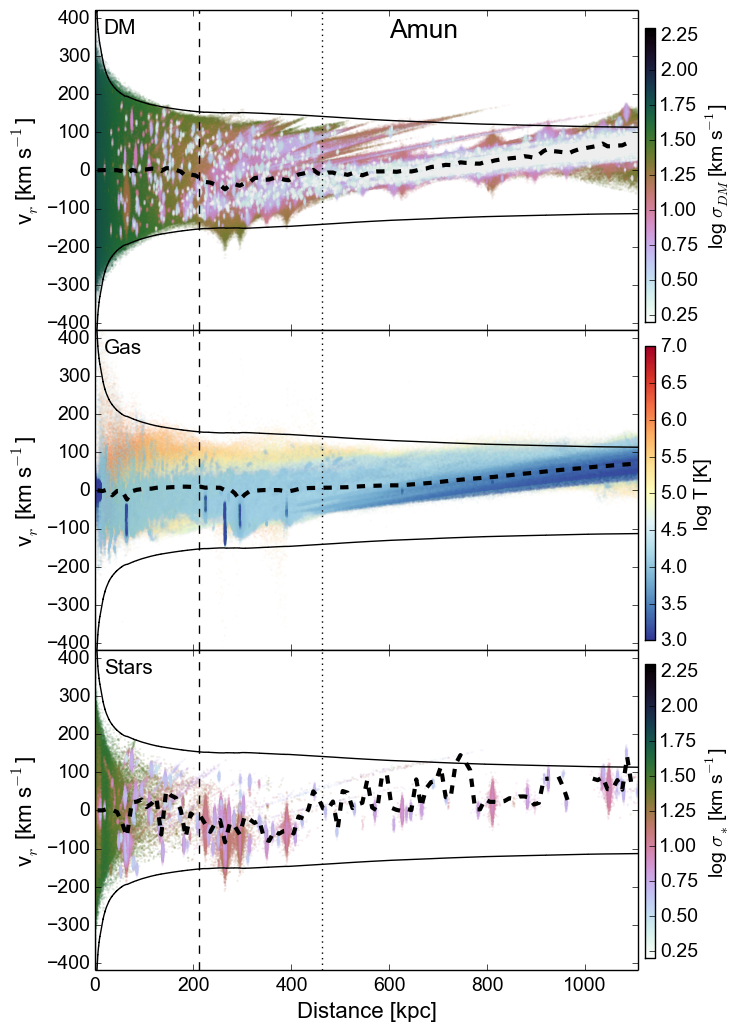}
  \caption{As in Fig.~\ref{fig:phase_space-s} but for Amun.}
  \label{fig:phase_space-a}
\end{figure}

The way in which (sub)structures and loose material accrete onto the
halo can be clearly seen in Figs.~\ref{fig:phase_space-s} and
\ref{fig:phase_space-a} (for Supay and Amun respectively, the diagrams
for Siris and Abu are very similar). These show the radial phase-space
distribution at $z=0$ for DM (top), gas (middle) and stars (bottom).
The dashed and dotted vertical lines indicate $\Rh$ and $\Rspl$,
respectively. The solid curves above and below the point distributions
are the escape velocity as a function of distance from the halo centre
while the thick dashed line represents the local mean radial velocity.
For DM, colour coding reflects the particle velocity dispersion
($\sigma_{\varv}$) calculated with a standard SPH-like kernel with 64
neighbours. We use $\sigma_{\varv}$ since it is a good indicator of
the kinematical state of the immediate neighbourhood around a given
particle. For example, a particle located in a virialized region
(i.e. where random motions dominate over coherent or orderly motions)
will have a large $\sigma_{\varv}$, while a particle moving within a
region where coherent motions dominate (i.e. filaments or voids of the
cosmic web) will have a lower $\sigma_{\varv}$.

Light colours denote substructures and the cosmic environment
(filaments) surrounding the haloes. Darker tones indicate the
potential well of the main objects. Intermediate hues show mixing
between these regions.  The needle-shaped features with positive
radial velocity ($\varv_r$) are tidal tails formed by the disruption
of substructures orbiting the haloes \citep[e.g.,][see also paper
I]{vogelsberger+09,erd-II}.  A-haloes show clear accretion flows,
i.e., material that falls in from cold environments of the cosmic web
(whitish band at $r > \Rspl$).  The radial velocity dispersion of the
infalling matter (vertical thickness at fixed radius in the
phase-space diagram) increases as particles move towards the central
regions and interarct with the halo and substructures.  This process
progressively enhances the random component of the particle
velocities, weakening the coherent accretion flow (white region on the
right-hand side of the phase-space diagram) that becomes less and less
noticeable while moving inwards (see also bottom panel of
Fig.~\ref{fig:dm_accretion}).  This effect becomes stronger as the
halo grows in mass and radius since the accretion flow needs to cross
larger an denser virialized regions.  Notice that S-haloes do not show
coherent infalling motions (see paper I). The radial-velocity
distribution is approximately symmetric between positive and negative
values until very large distances from the halo center.

\subsubsection{Gas accretion within haloes}
\label{sec:gas}

\begin{figure}
  \includegraphics[width=\columnwidth]{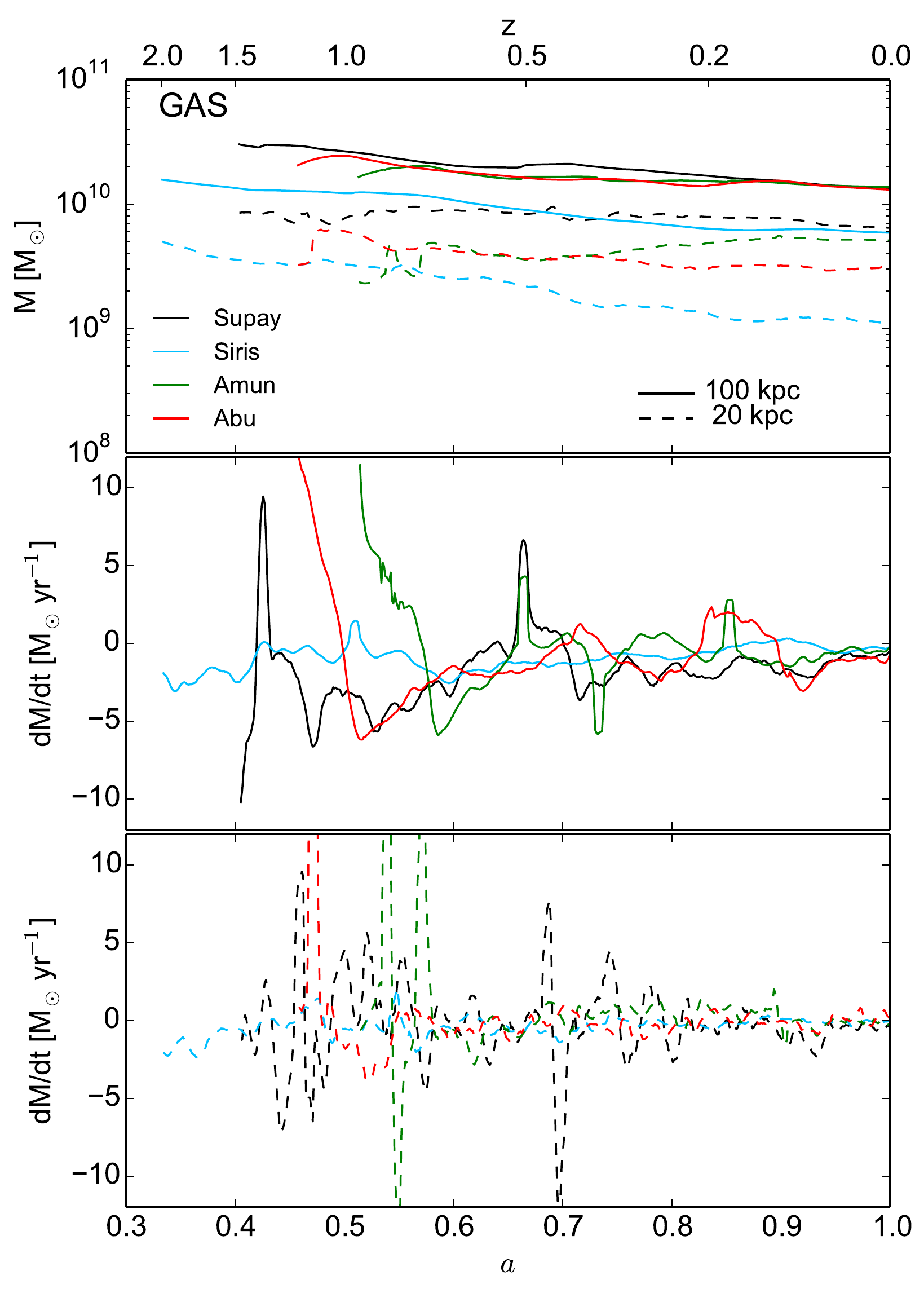}
  \caption{As in Fig.~\ref{fig:dm_r} but for the gas component. Notice
    that the vertical range and scale are different between the two
    figures.}
  \label{fig:gas_r}
\end{figure}

The gas component is subject to more physical processes than DM. Apart
from gravity, it is also severely affected by the combined action of
star formation, winds and feedback.  We have analysed gas accretion
onto the halo along the same lines as we did for the DM component
(Section~\ref{ss:dm}). In the top panel of Fig.~\ref{fig:gas_r}, we
show the gas mass within $r_{100}$ and $r_{20}$. A difference from the
DM case is that the gas content decrease at $r_{100}$ by a factor of
two for the S-haloes and 1.5 for the A-ones as time progresses.
Similarly, at $r_{20}$ the amount of gas slightly decreases, with the
exception of Amun in which remains constant at late times (see also
the bottom panel of Fig.~\ref{fig:gas_r}).

\begin{figure*}
  \includegraphics[width=7in]{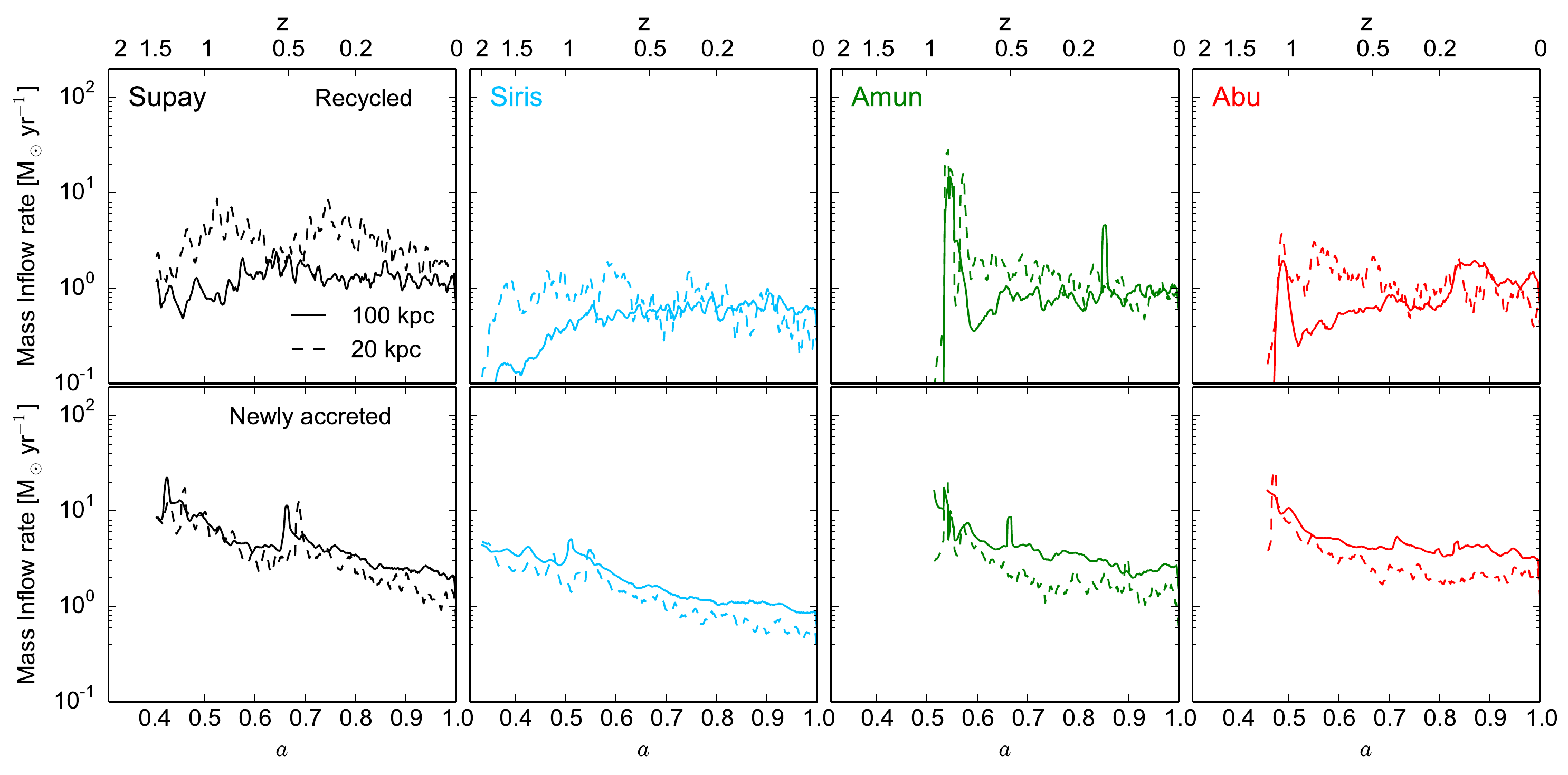}
  \caption{As in Fig.~\ref{fig:dm_accretion} but for the gas
    component. Notice that the vertical range and scale are different
    between the two figures.}
  \label{fig:gas_accretion}
\end{figure*}

We define recycled gas as the material that has been expelled from a
given region due to SNe, winds or tidal stripping from infalling
substructures.  As in the case of the DM, the recycled gas accretion
rates at $r_{100}$ increase up to $z>0.5$ after which they remain
almost constant for the S-haloes while in the A-ones still register a
slight increase (Fig.~\ref{fig:gas_accretion}, top panel). The
situation is different at $r_{20}$ where the effects due to the
presence of the central galaxy are much stronger. Although, the curves
are noisy due to the presence of incoming and outgoing substructures,
the main trends are clear: the recycled material decreases for the
S-haloes while for the A-ones remains almost constant.

The amount of newly accreted gas decreases at both radii with time
from $10~\msunyr$ down to $1-2~\msunyr$ for the S-haloes and down to
$3-4~\msunyr$ for the A-ones (Fig.~\ref{fig:gas_accretion}, bottom
panel).  Furthermore, the accretion rates for the A-haloes flatten at
both radii so that they are slightly higher than those from their
stalled counterparts. Overall, the fresh-gas accretion rates are in
agreement with \citet{brook+14} who found similar values for galaxies
of similar masses.  Local maxima (spikes) in the gas inflow rate are
much less prominent than in the DM. This is because gas that is
brought by infalling substructures is mostly stripped from them due to
ablation, shock heating and/or ram pressure. Therefore, most of the
substructures that cross the central region are gas naked (see also
paper III).  The total amount of gas that is being recycled is of the
same order as the newly accreted gas at late times. However, at early
times the trends are different, new gas accretion exceeds by almost an
order of magnitude the amount of recycled gas.

Radial gas motions are investigated in the middle panel of
Figs.~\ref{fig:phase_space-s} and \ref{fig:phase_space-a}, where the
colour coding reflects the gas temperature $T$.  At distances larger
than $\Rspl$, a clear accretion pattern is noticeable both in Supay
and Amun: a coherent stream of cold gas (blue) with negative radial
velocity (the same holds true for the remaining two haloes).  [A
relatively massive halo is infalling onto Supay together with its own
gaseous accretion stream at $r \approx 530~$kpc.]  It is worth
mentioning that, in the DM, this feature was only visible for the
A-haloes (see the top panels). Infalling cold gas gradually gets
warmer as it approaches the halo boundaries.  In the surroundings of
the haloes, there is also a warmer gas component (yellow) generated by
the combined action of SF and feedback from substructures as well as
accretion shocks due to large density gradients in the cosmic web.
Hotter gas within $\Rh$ has been shock heated as it fell into the halo
(orange). Islands of very cold gas (dark blue) identify the locations
of galaxies which are associated with massive DM substructures, while
smaller subhaloes are devoid of gas.  Feedback processes increase the
gas temperature and provide sufficient kinetic energy to expel gas
from its host halo.  As the gas moves outwards, it radiates energy
away, cools down and, if still gravitationally bound, reduces its
radial velocity until it reaches $\varv_r = 0\kms$. After this, the
material slowly falls back into the halo due to its gravitational
pull, increasing again the gas temperature. Unbound gas moves quickly
away from the halo.

\begin{figure}
  \includegraphics[width=\columnwidth]{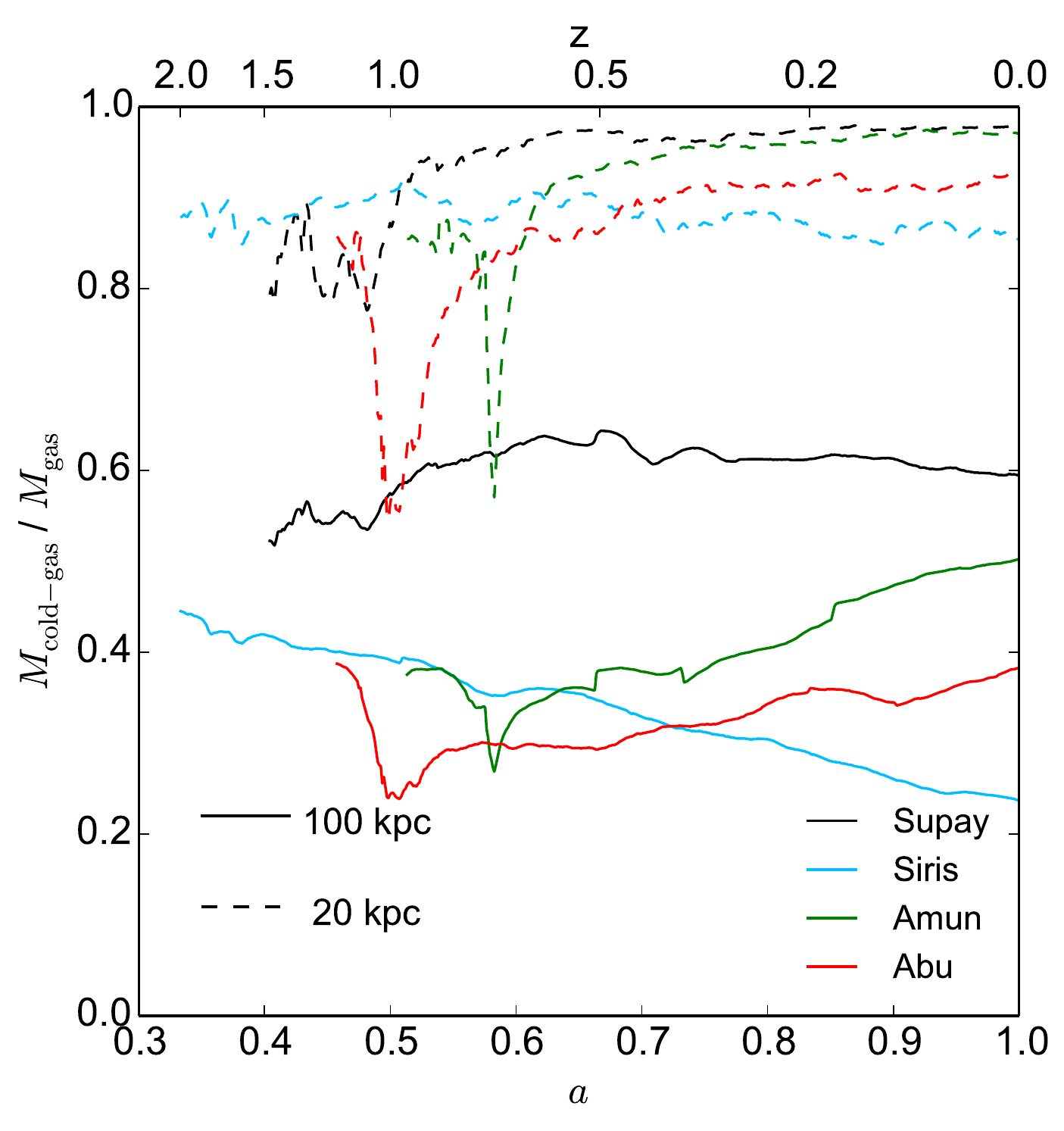}
  \caption{Evolution of the cold-gas ($T <3\times 10^4$K) mass
    fractions for our galaxy set. The solid lines indicate the
    mass fractions calculated at $r_{100}$ while the dashed lines at
    $r_{20}$.}
  \label{fig:cold-gas}
\end{figure}

An important quantity is the amount of cold gas ($T \leq 3\times
10^4$K) that might fall onto the central galaxy and support SF
(further details will be discussed in Section~\ref{sec:gals}). In
Fig.~\ref{fig:cold-gas} we have calculated the evolution of the
cold-gas fraction at $r_{100}$ (solid) and $r_{20}$ (dashed) for our
haloes. The inner regions are dominated by the cold component while in
the outer ones the warmer gas is equally important.  In terms of sheer
numbers all haloes present similar fractions, however, trends in their
time evolution within $r_{100}$ are different between A- and
S-objects. The former show a constant increase while the latter do not
(Supay has no change and Siris a constant decrease).  Instead, in the
central regions, all haloes have similar and little evolving cold-gas
fractions.  All this suggests that the amount of cold gas in the halo
cores does not correlate with the halo assembly history.  We explain
this as follows.  Gas that has been shock heated when it enters the
halo (or pushed out by feedback) cools down radiatively and eventually
rains into the central part.  Since cooling times are long (comparable
with the Hubble time, i.e., several Gyr), the halo is filled with
large reservoirs of hot gas. Therefore, if gas accretion in the
outskirts would cease (which is not the case even for S-haloes), cold
gas would continue infalling onto the central galaxy for very long
time.

In summary, haloes of the same mass but with extremely different
assembly histories do not show large differences in their central
cold-gas content if they are fed via the `hot-mode accretion'.
However, we find that a factor of two difference in the amount of
infalling gas per unit time should be expected with the A-haloes
showing the highest rate.

\subsection{The central galaxies}
\label{sec:gals}

\begin{table*}
  \centering
  \caption{Galaxy properties at $z=0$. The first column denotes the model name inherited
    from their respective parent halo. The second and third columns indicate the
    galaxy stellar and gaseous masses. The fourth column presents the S\'ersic index of the stellar bulge.
    The fifth, sixth and seventh columns the stellar
    bulge, disc, and vertical scale-lengths respectively. The eighth column shows
    the stellar disc-to-total mass ratio. The ninth column represents 
    the slope of the metallicity gradient of the stellar disc. The age of the stellar disc is 
    indicated in column ten. The Petrosian radius is shown in the eleventh 
    column and the CAS statistics (Concentration, Asymmetry and Clumpiness) 
    inferred from the stellar distributions in the last three columns.}
  \label{tab:galaxies}
  \begin{tabular}{lccccccccccccc}
    \hline
    Galaxy name & $M_*$ & $M_{\rm gas}$ & $n_{\rm b}$ &$R_{\rm b}$& $R_{\rm d}$ & $z_{\rm h}$ & D/T & m$_{\rm Z}$ & $t_{\rm d}$ & $R_{\rm petros}$ & C & A & S \\
                   & [$10^{9} \msun$] & [$10^{9} \msun$] & & [$\kpc$] & [$\kpc$] & [$\kpc$] & & [${\rm dex}\kpc^{-1}$] & [Gyr] & [$\kpc$] & & & \\
    \hline
    Supay & 23.9 & 6.57 & 0.63 & 1.65 & 3.20 & 0.60 & 0.69 &-0.01 & 10.5 & 12.75 & 4.37 & 0.17 & 0.25 \\
    Siris & 11.4 & 1.07 & 0.90 & 2.46 & 0.64 & 0.40 & 0.41 &-0.04 & 11.2 & 10.77 & 3.04 & 0.27 & 0.28 \\
    Abu   & 20.3 & 3.31 & 0.66 & 3.12 & 0.54 & 0.20 & 0.43 &-0.06 &  6.8 & 10.72 & 3.13 & 0.28 & 0.22 \\
    Amun  & 12.1 & 5.16 & 0.89 & 1.27 & 2.28 & 0.28 & 0.66 &-0.02 & 7.1 &11.72 & 3.55 & 0.21 & 0.32 \\
    \hline
  \end{tabular}
\end{table*}

\begin{figure*}
  \subfloat{\includegraphics[width=1.7in]{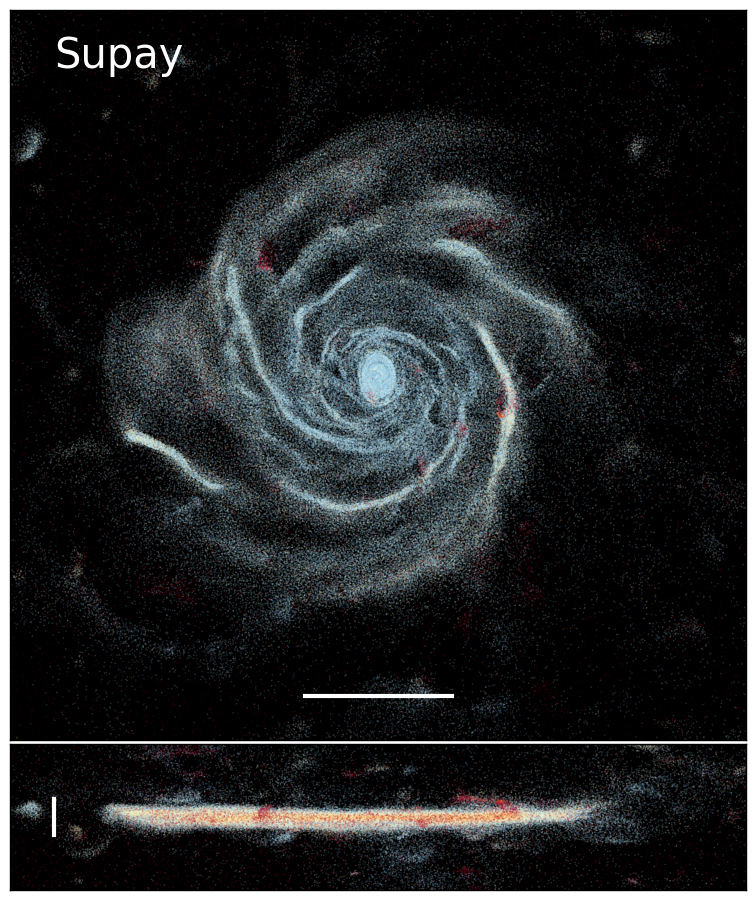}}
  \subfloat{\includegraphics[width=1.7in]{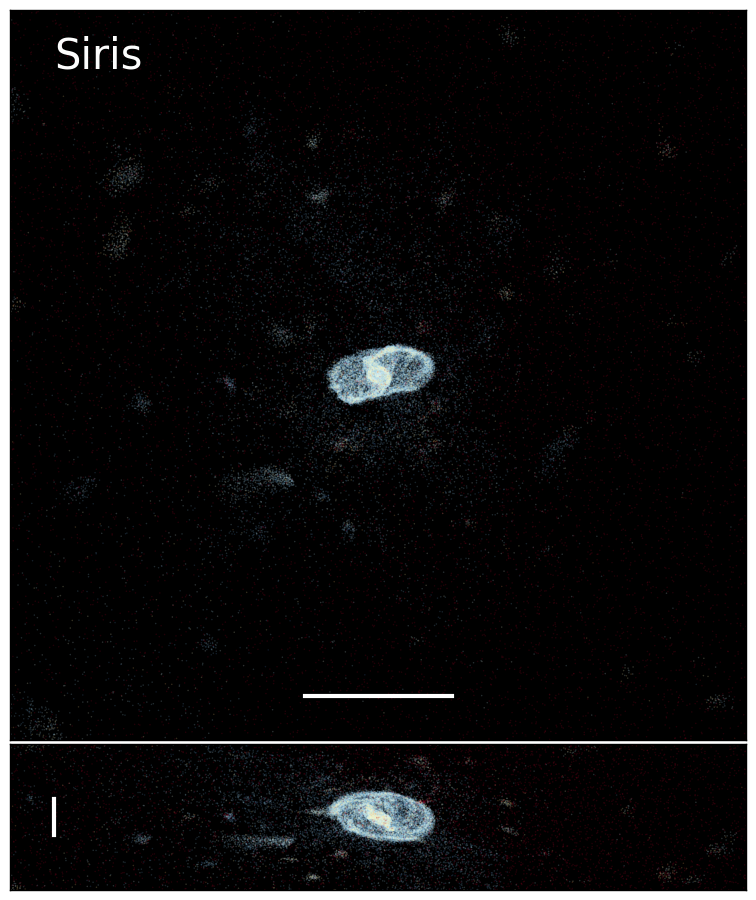}}
  \subfloat{\includegraphics[width=1.7in]{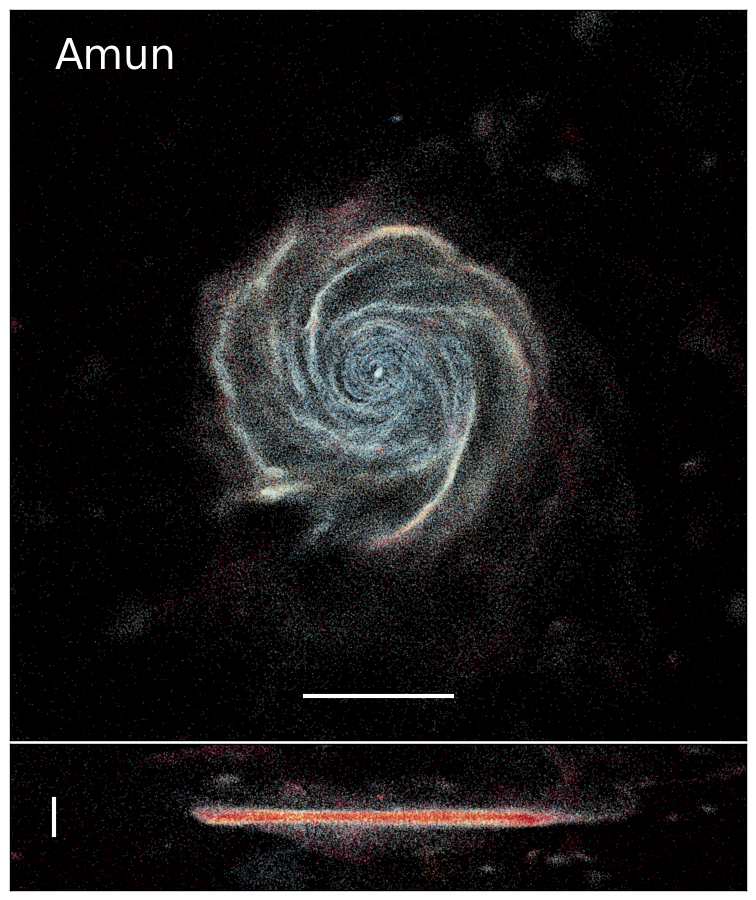}}
  \subfloat{\includegraphics[width=1.98in]{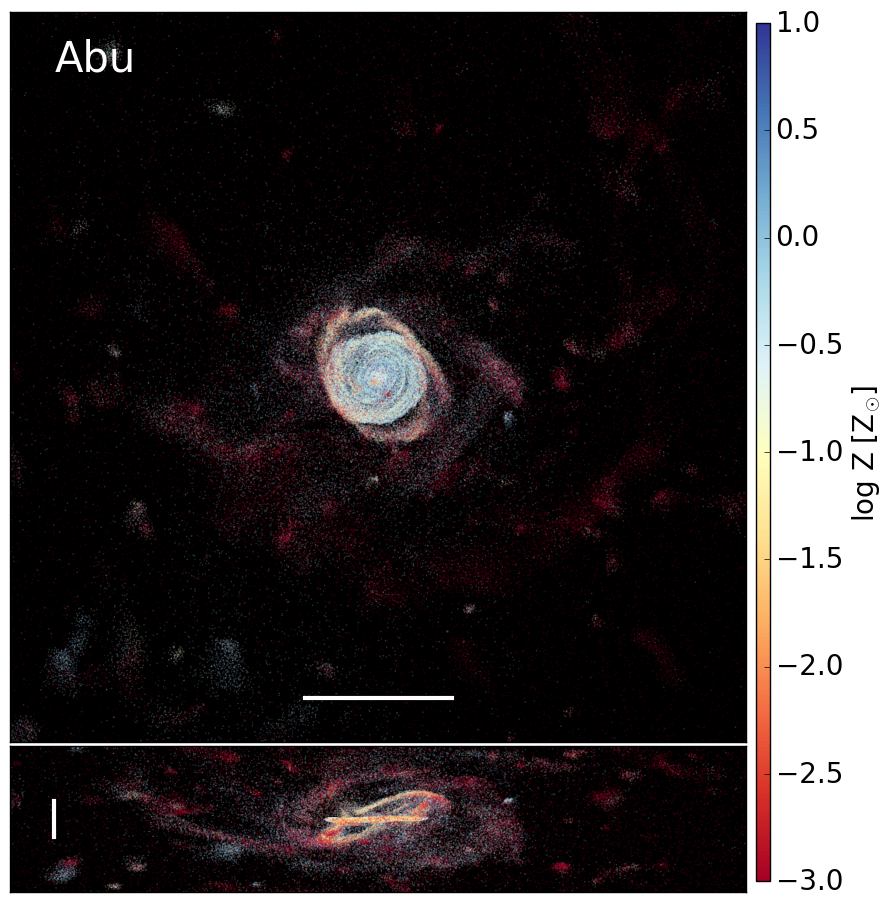}}
  \caption{Gas content of central galaxies at $z=0$.  The face-on
    (top) and edge-on (bottom) images have been obtained using the
    direction of the stellar angular momentum.  Colour coding maps the
    local metallicity $Z$. The white bars are yardsticks of 10
    (horizontal) and 5 (vertical) kpc.}
  \label{fig:galaxies-gas}
\end{figure*}

\begin{figure*}
  \subfloat{\includegraphics[width=1.7in]{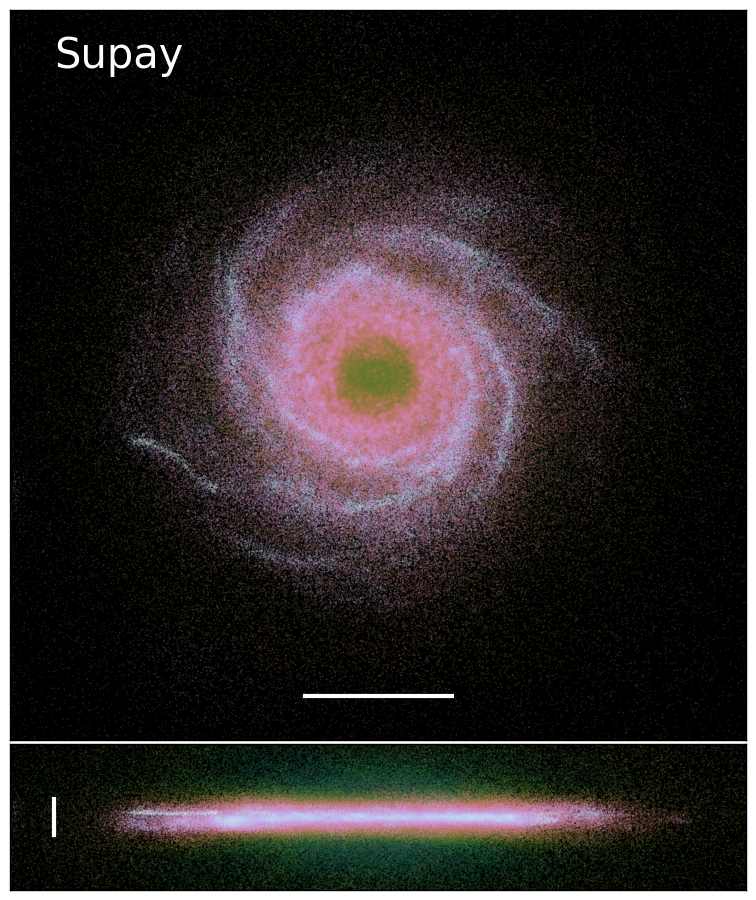}}
  \subfloat{\includegraphics[width=1.7in]{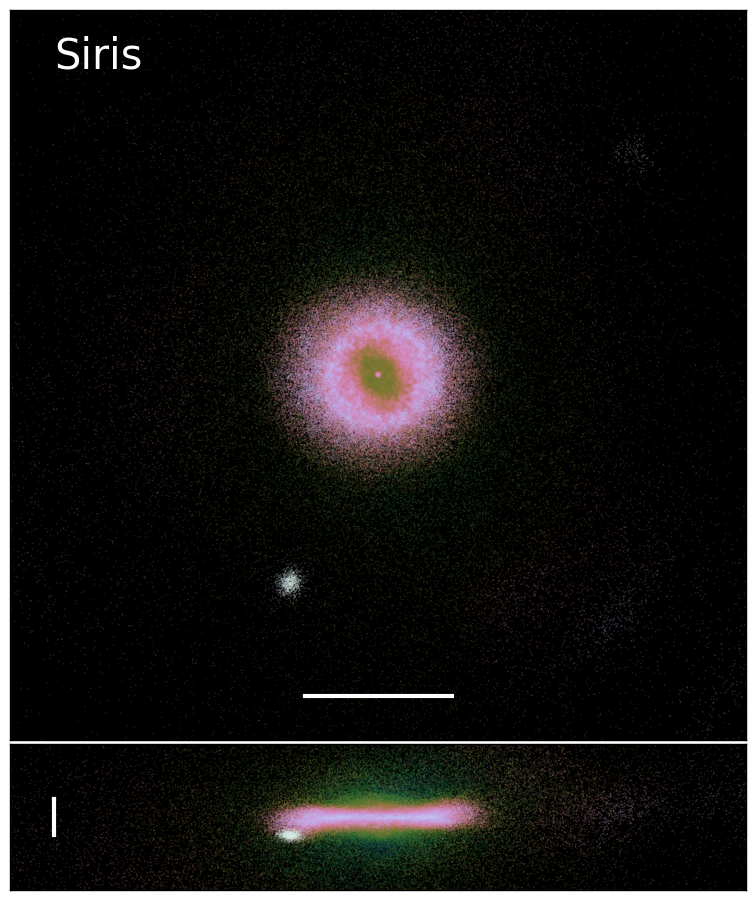}}
  \subfloat{\includegraphics[width=1.7in]{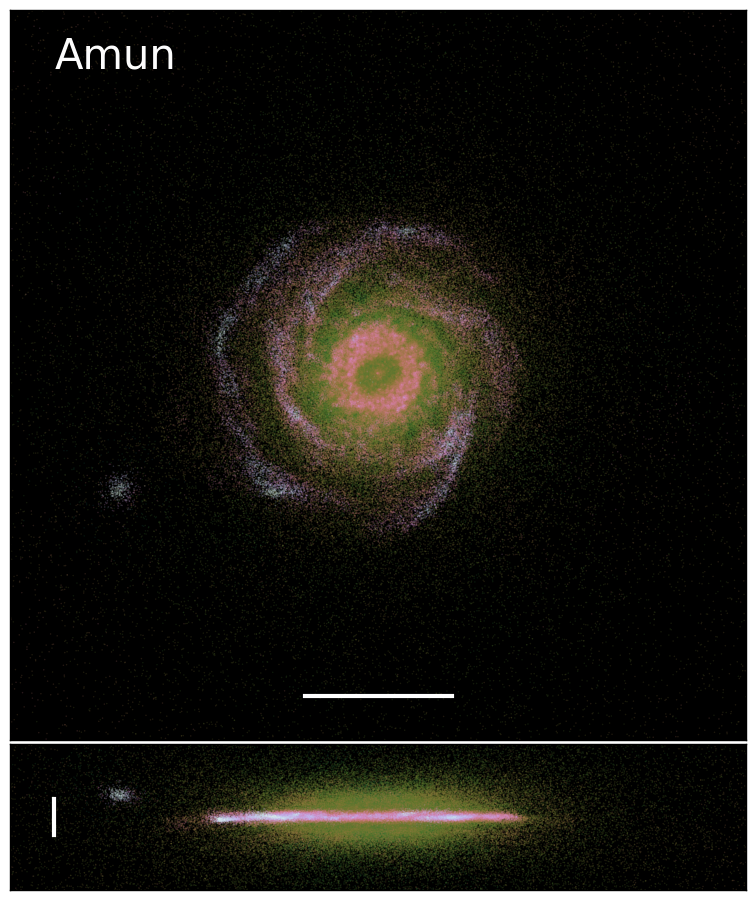}}
  \subfloat{\includegraphics[width=1.98in]{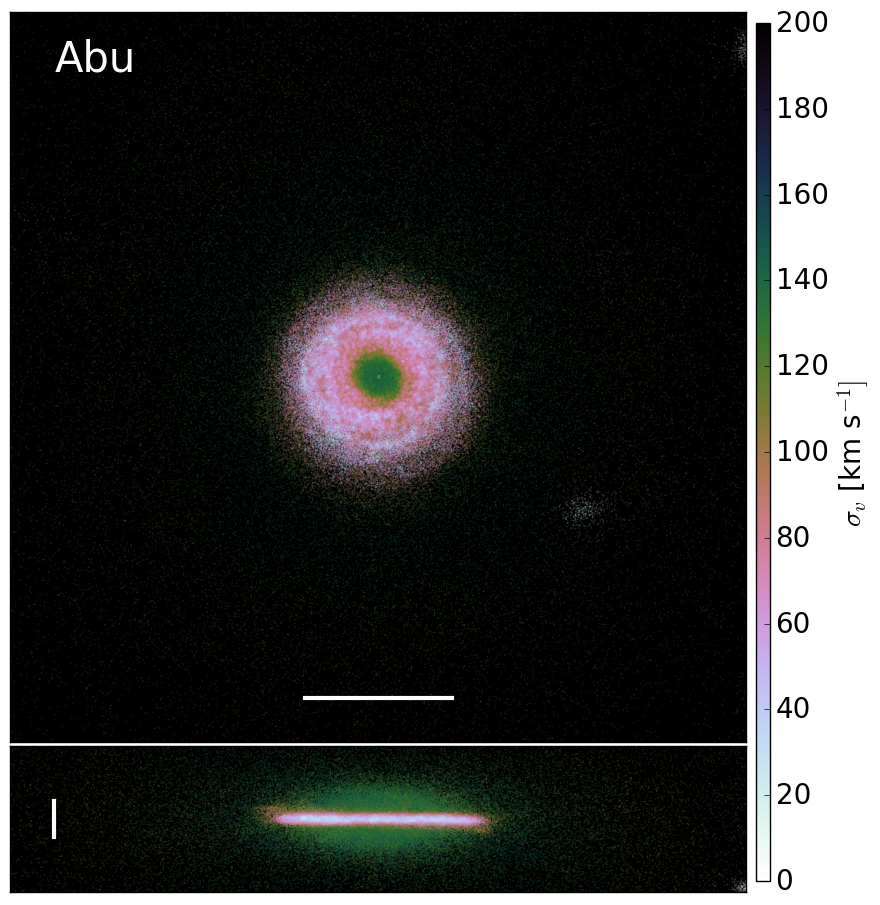}}
  \caption{As in Fig.~\ref{fig:galaxies-gas} but for the stellar
    component colour coded with the local velocity dispersion
    $\sigma_{\varv}$.}
\label{fig:galaxies-stars}
\end{figure*}

We now turn our attention to the central galaxies within our halo
sample.  We will refer to the galaxies by using the same names of their
respective host haloes.  Figs.~\ref{fig:galaxies-gas} and
\ref{fig:galaxies-stars} show the face-on (top) 
and edge-on (bottom) gas and stellar distributions at $z=0$,
respectively. Colour coding reflects the metallicity ($Z$) for the gas
and the local velocity dispersion ($\sigma_{\varv}$) for the stars.
The white horizontal and vertical segments represent 10 and 5 kpc
yardsticks, respectively.  Based on their morphologies, our galaxies
can be divided into two groups that, however, do not separate early
and late collapsing haloes.  While Supay and Amun are grand design
spirals with thin gaseous discs, Siris and Abu are smaller
bulge-dominated disc galaxies with more irregular gas distributions.
In particular, Abu has been perturbed by minor mergers at $z \approx
0.2$ as indicated by: (i) the lack of alignment between the gaseous
and stellar discs; (ii) the presence of a ring-like pattern around the
gaseous disc with a different metallicity.  These results are based on
visual inspections of the figures, a more quantitative study will be
addressed in Section~\ref{sec:decomposition}.

\begin{figure}
  \includegraphics[width=\columnwidth]{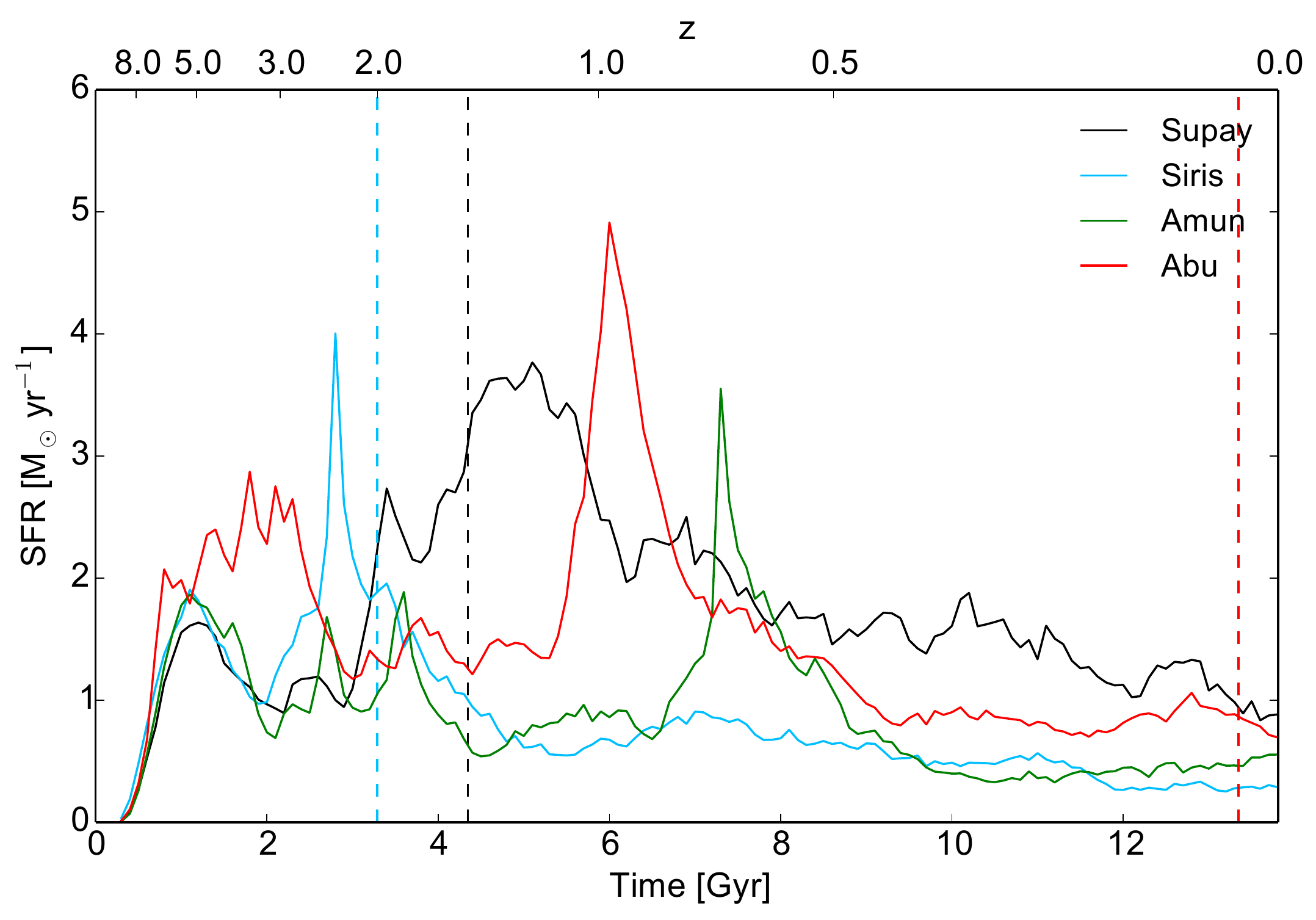}
  \caption{SF histories for our central galaxies. The dashed-vertical
    lines indicate the halo collapse times (notice that for Amun $\ac
    > 1$).}
  \label{fig:sfr}
\end{figure}

Fig.~\ref{fig:sfr} shows the SF histories for our galaxies.  The
highest peak for each model coincides with the last major galaxy
merger.  A-haloes experience this at lower redshifts ($z \lesssim 1$)
with respect to the S-ones.  For the latter class, the peak also
  coincides with the halo collapse time (dashed-vertical lines).  At
later times the SFR decreases to $1~\msunyr$ or less.

Table~\ref{tab:galaxies} summarizes the main physical properties of
the galaxies (see the remainder of Section~\ref{sec:results} for the
precise definitions).

\subsection{Comparison with observations}
\label{sec:observations}

\begin{figure}
  \includegraphics[width=\columnwidth]{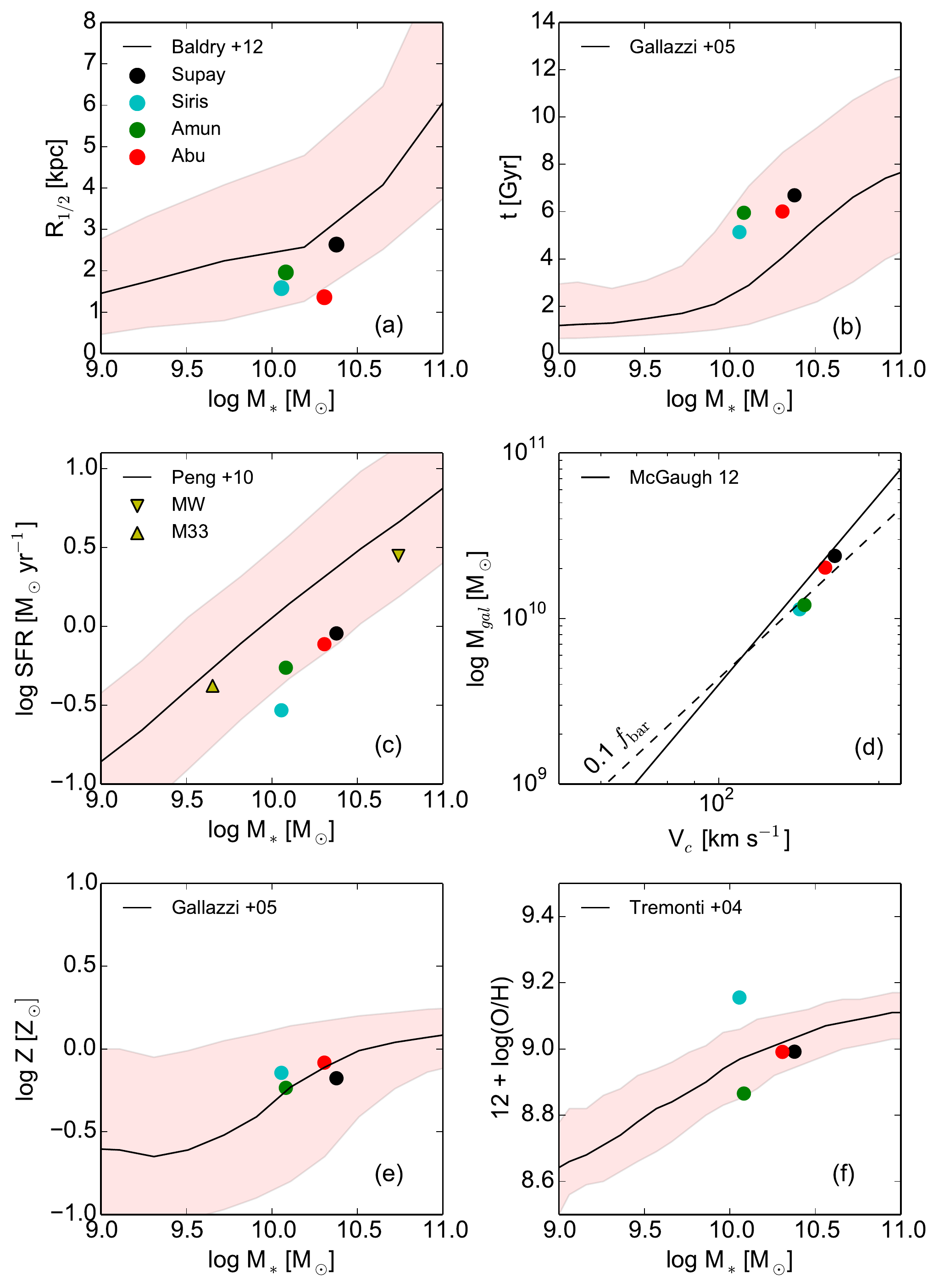}
  \caption{Simulated galaxies vs observations at $z=0$. In each panel
    individual galaxies are indicated by the same colour, the shaded
    regions represent the $1\sigma$ dispersion around the respective
    observational mean values (black lines). a) Half-mass radius vs
    stellar mass. b) Mean stellar age vs stellar mass. c) Star
    formation rate vs stellar mass (the galaxy main sequence), the
    triangles indicates the values for the MW and M33. d) Baryonic
    Tully-Fisher relation. e) Stellar metallicity vs stellar mass. f)
    Gas metallicity abundances vs stellar mass.}
    \label{fig:gal-properties}
\end{figure}

In order to assess the reliability of our simulated galaxies, we have
compared our suite to a set of observational properties at $z=0$
gathered from the literature.
Fig. ~\ref{fig:gal-properties} presents the following data: 
(a) the stellar mass -- half mass radius relation \citep{baldry+12};
(b) the age -- stellar mass relation \citep{gallazzi+05}; 
(c) the galaxy main sequence \citep{peng+10};
(d) the baryonic Tully-Fisher relation \citep{McGaugh12};
(e) the stellar metallicity -- stellar mass relation \citep{gallazzi+05};
(f) the gas metallicity -- stellar mass relation \citep{tremonti+04}.  
In all cases the black lines represent the mean of the measured
values, while the shaded regions indicate the uncertainties
($\pm1 \sigma$). The simulations are represented by filled circles
with different colours. In panel (c) we have added the observed values
for the Milky-Way and M33 that bracket our galaxy sample in terms of
stellar mass.

In general our galaxies are in very good agreement with the different
observational properties. The objects do not only follow the overall
trends, but they are also close to the mean values. If anything, our
simulations consistently produce slightly smaller, older and less star
forming galaxies than the observed mean values but lie well within the
scatter.

\subsection{Galactic decomposition}
\label{sec:decomposition}

In order to characterise the stellar morphology of our galaxy set we
have implemented two different and complementary methods: fitting
their stellar surface density profiles and using a kinematical
decomposition.

\subsubsection{Surface density profiles}
\label{sec:gal-profiles}

\begin{figure}
  \includegraphics[width=\columnwidth]{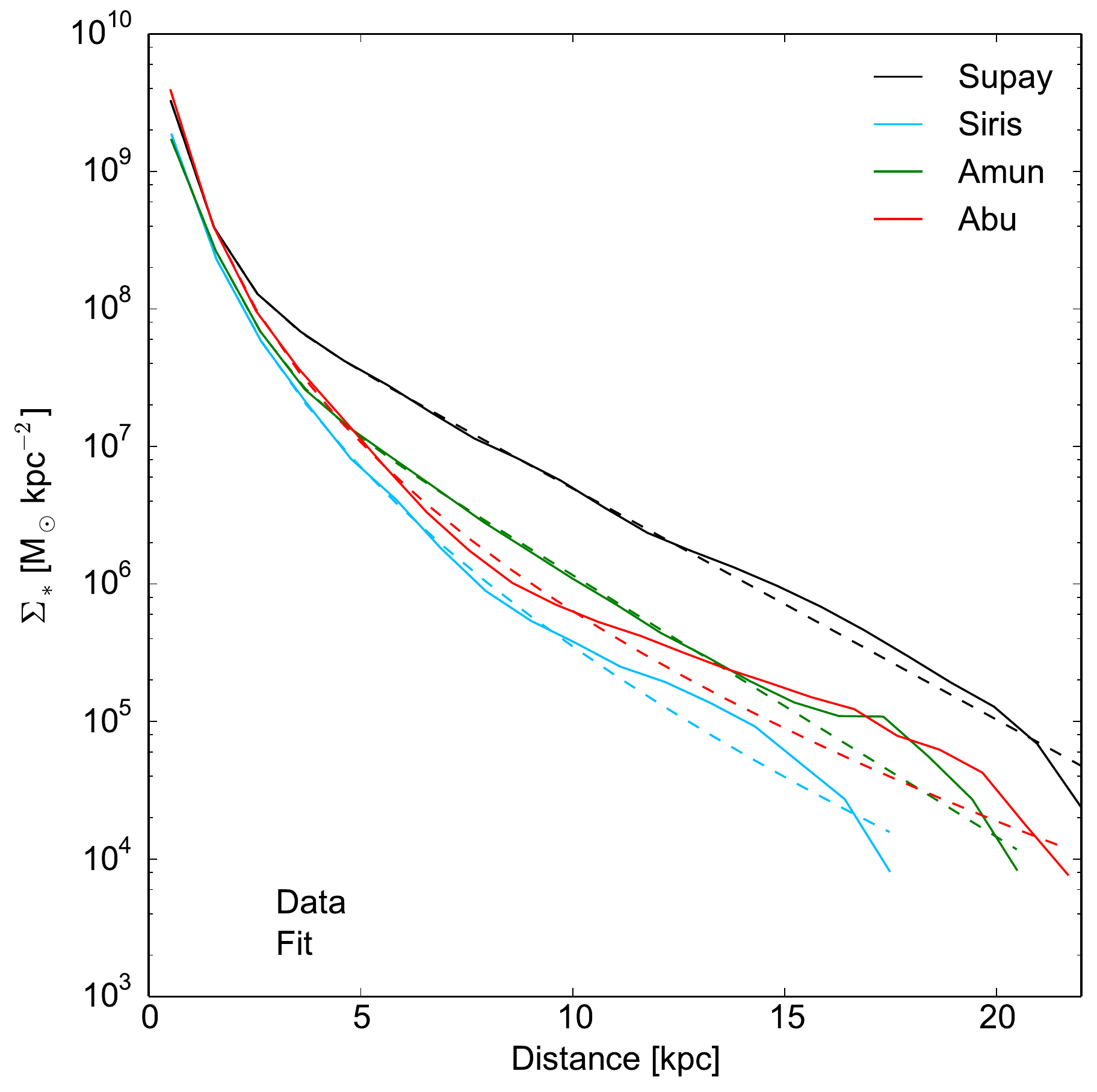}
  \caption{Stellar surface density profiles. The profiles have been
    calculated for face-on galaxy projections, indicated by the
    solid lines. The dashed lines represent the best-fitted model
    consisting of a bulge (S\'ersic profile) and a disc (exponential
    disc).}
    \label{fig:surf-prof}
\end{figure}

Fig.~\ref{fig:surf-prof} shows the azimuthally-averaged face-on
surface density profiles (solid lines). We have fitted the data with
the following model:
\begin{equation}
	\Sigma(r) = \Sigma_{\rm b} \exp \Bigg\{-b_n \bigg[ \bigg(\frac{r}{R_{\rm b}}\bigg)^{1/n_{\rm b}} -1 \bigg] \Bigg\} + \Sigma_{\rm d} \exp \Bigg( -\frac{r}{R_{\rm d}} \Bigg) \,,
 	\label{eq:surf_prof}
\end{equation}
which includes two components, a bulge\footnote{For simplicity we have
  included the central bar component into the contribution of the
  galactic bulge.} and an exponential disc.  The first term on the
right-hand side, equivalent to a standard S\'ersic profile
\citep{sersic63}, has been expressed in terms of effective quantities
to facilitate its comparison with observations
\citep{macarthur+03}. In this way, $R_{\rm b}$ is the effective bulge
radius, $\Sigma_{\rm b}$ the density at this point and $n_{\rm b}$ the
S\'ersic index. Note that $b_n$ depends on $n_{\rm b}$
\citep{ciotti-bertin,macarthur+03}.  The disc component is
parameterised in terms of its scale length $R_{\rm d}$ and the density
$\Sigma_{\rm d}$.  The fits were performed by simultaneously varying
the parameters.  Fitting the disc first and then the bulge gave
consistent results ~\citep[see also][]{marinacci+14}. The dashed lines
in Fig~\ref{fig:surf-prof} represent the best fits to the data, the
corresponding scale lengths are reported in Table~\ref{tab:galaxies}.
These results corroborate the visual impression described in
Section~\ref{sec:gals}, namely, Supay and Amun have extended discs,
while Abu and Siris more compact ones.

We have calculated the bulge and disc masses by integrating
equation~(\ref{eq:surf_prof}). Table~\ref{tab:galaxies} lists the
disc-to-total mass ratio (D/T) for our sample.  This indicates that
Supay and Amun are disc dominated. On the other hand, Siris and Abu
have the most dominant bulges of the sample.

\subsubsection{Kinematic decomposition}
\label{sec:kdecomposition}

\begin{figure}
  \includegraphics[width=\columnwidth]{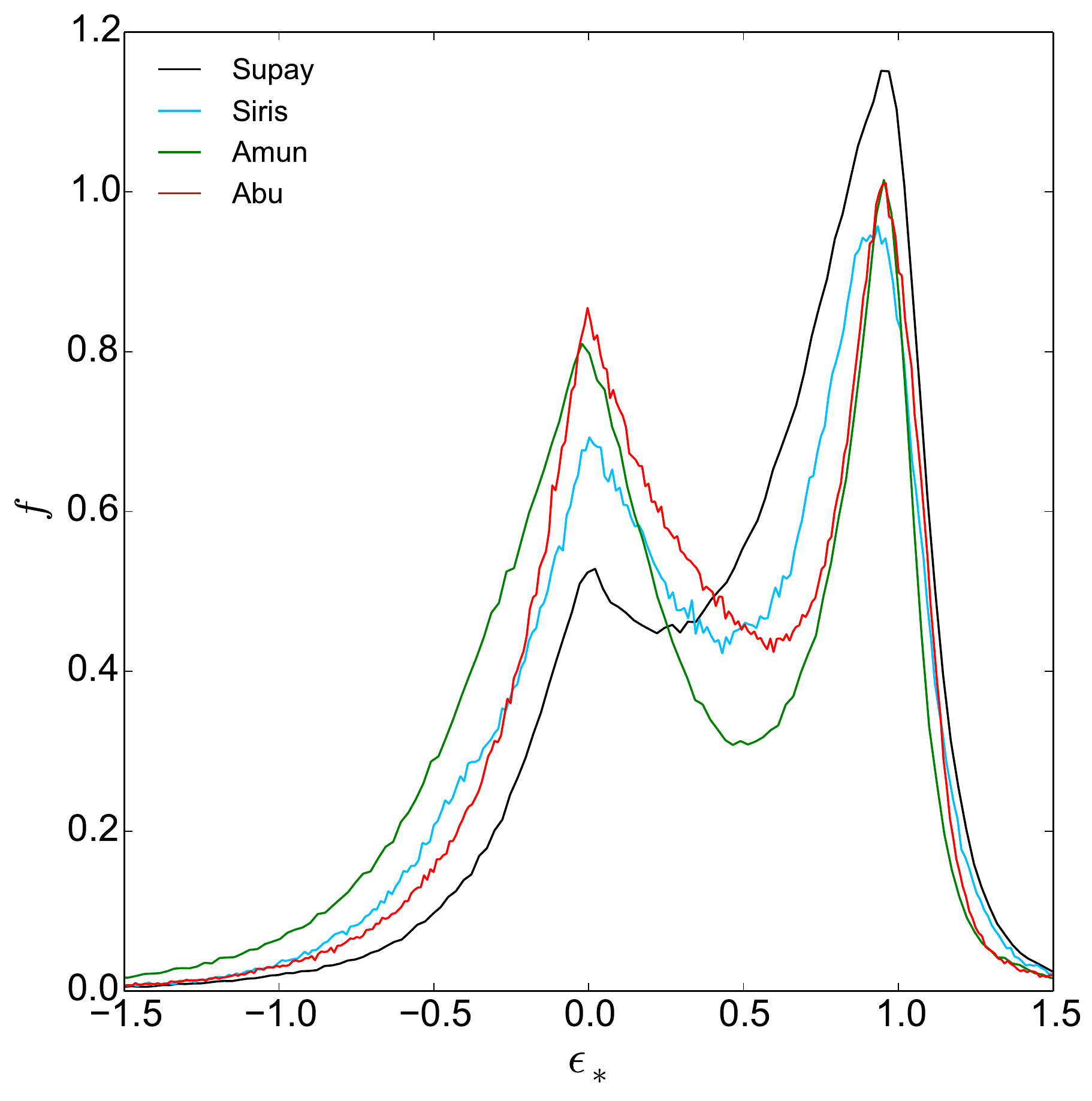}	
  \caption{Distribution of the stellar circularity parameter
    $\epsilon_*$ at $z=0$. Different colours represent different
    galaxies. The peak around 0 indicates the presence of a bulge
    component, while the peak at 1 the presence of a rotationally
    dominated disc component.}
\label{fig:circularity}
\end{figure}

We performed a kinematic bulge-to-disc decomposition in
terms of the circularity parameter 
\begin{equation}
	\epsilon_* = \frac{j_{\rm z}}{j_{\rm circ}} \,,
\end{equation}
\citep{abadi+03,scannapieco+08} where $j_{\rm z}$ is the component of
the specific angular momentum of each particle parallel to the total
stellar angular momentum and $j_{\rm circ} = r V_{\rm circ}$, with
$V_{\rm circ}^2 = GM(<r)/r$. Fig.~\ref{fig:circularity} shows the
distributions of $\epsilon_*$ for our four galaxies normalised so that
$\int f(\epsilon_*)\,{\rm d}\epsilon_* =1$. The histograms extend
beyond $\epsilon_* = \pm1$ because we also considered particles that
are not gravitationally bound to the galaxies.  Particles in
dispersion-dominated regions (bulges) have $\epsilon_* \sim 0$, while
those in rotation-supported ones (discs) $\epsilon_* \sim 1$.  We have
calculated the masses of the disc and bulge components. The former is
defined by stars with $\epsilon_* > 0.7$ \citep[e.g.,][]{marinacci+14}
and imposing an extra condition of $\sigma_{\varv} \leq 50\kms$.  The
remaining particles are very extended spatially, in contrast with the
bulge definition used in Section~\ref{sec:gal-profiles}.  Therefore,
we have made use of the galactic rotation curves in order to separate
this spheroidal component into a bulge and a halo. Namely, we
considered as bulge-particles those with $\epsilon_* < 0.7$ and
$r<2r_{\rm max}$, with $r_{\rm max}$ the radius of the maximum of the
rotation curve. The halo is composed by the remaining particles.  The
resulting D/T values are in very good agreement (within 10 per cent)
with those reported in Table~\ref{tab:galaxies} obtained from the
stellar surface density profiles. Overall, our values are on the
low-side of the observed range in the CALIFA survey
\citep{mendez-abreu+16}, in agreement with other numerical simulations
\citep{brooks-christensen16}. Anyway, Supay and Amun are within one
standard deviation from the CALIFA mean value.

\begin{figure}
  \includegraphics[width=\columnwidth]{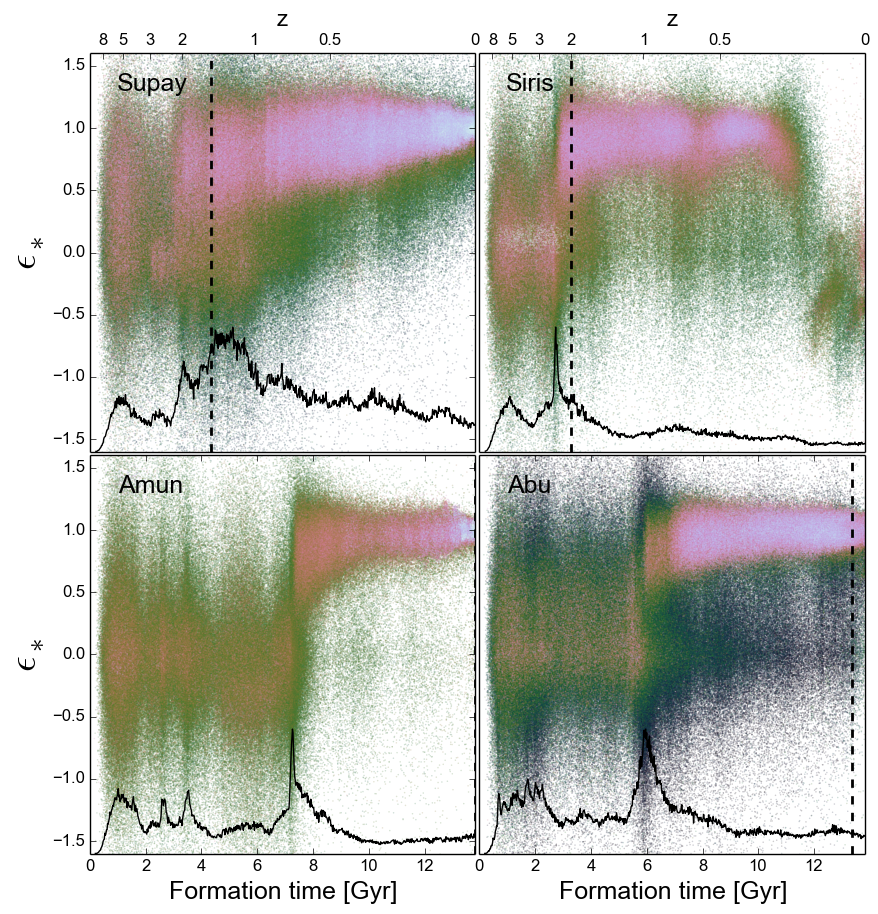}
  \caption{Circularity parameter calculated at $z=0$ as a function of
    the stellar age expressed in terms of the formation time. The
    colour-coding is proportional to the local stellar velocity
    dispersion $\sigma_{\varv}$. The vertical spikes indicate bursts
    in the SFR (represented by the bottom curves in arbitrary units)
    mostly influenced by merger events. The dashed-vertical lines
    indicate the halo collapse times.}
\label{fig:circularity-t}
\end{figure}

In Fig.~\ref{fig:circularity-t} we have plotted $\epsilon_*$
(evaluated at $z=0$) as a function of the stellar formation time
\citep[see also][]{sales+12,aumer+13}.  Colour coding represents the
local stellar velocity dispersions $\sigma_{\varv}$. Generally, stars
that now make the spheroidal components formed first while those found
in the disc at later times. All galaxies show a sharp transition after
which the stellar disc forms. We refer to this event as the disc
formation time (see Table~\ref{tab:galaxies}). This happens early on
for Siris and Supay ($z \approx 2-3$) and approximately coincides with
the halo formation time (dashed-vertical lines) while it is delayed
until $z < 1$ for Abu and Amun.  The transition epochs coincide with
the maximum of the SF histories for all galaxies (represented by the
black-solid curves at the bottom).  We conclude that the age of the
stellar discs correlates with the collapse time of haloes in which the
galaxies are found.

\subsubsection{Vertical profiles}

\begin{figure}
  \includegraphics[width=\columnwidth]{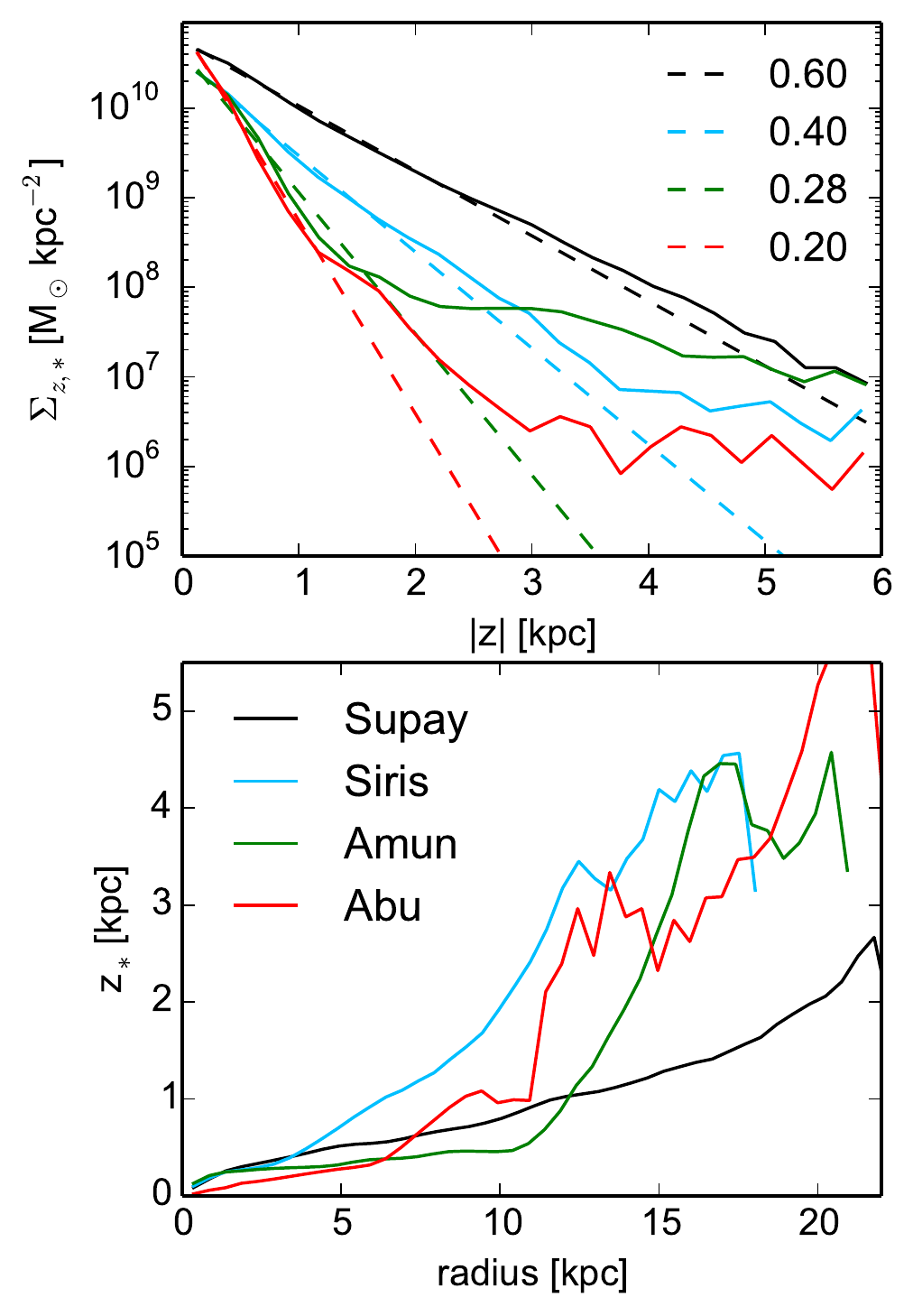}
  \caption{The top panel shows the stellar vertical density profiles
    (solid) with their corresponding best exponential fits (dashed).
    The quoted numbers represent the scale-heights, expressed in kpc.
    The bottom panel displays the mass-weighted rms value of the
    $z$-coordinate as a function of radius.}
  \label{fig:gal_vprof}
\end{figure}

The edge-on stellar distributions of Fig.~\ref{fig:galaxies-stars}
show a clear difference between the thickness of the galactic disc in
stalled and accreting haloes.  The former have a much thicker vertical
distributions than the latter.  In order to accurately measure this,
we fit the vertical density profiles with the relation
\citep{vanderkruit88}:
\begin{equation}
  \Sigma_{\rm z,*} = \Sigma_{\rm h} \exp \left(-\frac{z}{z_{\rm h}}\right) \,,
  \label{eq:s-vertical}
\end{equation}
where $z_{\rm h}$ is the vertical scale-height of the disc and
$\Sigma_{\rm h}$ is the midplane surface density. The top panel in
Fig.~\ref{fig:gal_vprof} shows the measured profiles (solid) and their
respective best fits (dashed). The corresponding values of $z_{\rm h}$
(in kpc) are also indicated in the figure and listed in
Table~\ref{tab:galaxies}. The average scale-height of the S-galaxies
is a factor of two larger than for the A-galaxies.

Two effects contribute to the thickening of the stellar vertical
profiles: i) (minor) mergers and close flybys
\citep[e.g.,][]{villalobos08,kazantzidis+08,purcell+10,zolotov+10,qu+11},
which give rise to flared discs as well as ii) gravitational
instabilities, which generate radially constant scale heights
\citep{bournaud+09}.  The bottom panel of Fig.~\ref{fig:gal_vprof}
shows that $z_*$, the rms value of the $z$-coordinate, increases as a
function of radius. This indicates that merger activity is probably
the main driver in the thickening of the discs \citep[see
however,][and references therein]{minchev+15}.

As we have shown in Fig.~\ref{fig:circularity-t}, S-galaxies have
older discs and therefore are more prone to experience mergers, flybys
of substructures and secular evolution. As already disccused in
Section~\ref{sec:gas},
most of the incoming
satellites are being stripped of their baryons and do not deposite gas
into the disc. A detailed analysis of the evolution and influence of
substructures will be discussed in paper III.

\subsection{CAS analysis}
\label{sec:cas}

We have looked for galactic signatures that could correlate with the
halo formation time. For this reason, we have measured galaxy
morphologies using the CAS (concentration, asymmetry, clumpiness)
non-parametric method \citep{conselice03,conselice14} applied to the
face-on stellar surface density. These quantities are designed to
capture the major features of the underlying galactic structures and
their evolution.  Concentration \citep[which correlates with the
S\'ersic index,][]{conselice03} is defined as:
\begin{equation}
  C = 5 \log_{10} \frac{r_{80}}{r_{20}} \,,
\end{equation}
where $r_{80}$ and $r_{20}$ are the circular apertures containing 80
and 20 per cent of the stellar mass enclosed within 1.5 Petrosian
radii.\footnote{In this context the Petrosian radius is defined as the
  location where the surface density divided by its average value
  within that radius reaches $\eta(R_{\rm petros}) = 0.2$ \citep{petrosian76}.}
Asymmetry is defined as (in the absence of a sky background):
\begin{equation}
  A = \frac{\sum | I - I_{180} |}{\sum I} \,,
\end{equation}
where $I$ and $I_{180}$ represent the galaxy map in its original form
and rotated by $180^\circ$, respectively. The sum is performed over
all pixels within 1.5 Petrosian radii.
Clumpiness (or smoothness) is defined as (in the absence of a sky background):
\begin{equation}
  S = \frac{\sum (I - B)}{\sum I} \,,
\end{equation}
where $I$ is defined as before and $B$ is the smoothed map. The scale
of the smoothing kernel ($\sigma$) depends on the radius of the galaxy
and is defined as $\sigma = 0.3 R_{\rm petros}$.

The results reported in Table~\ref{tab:galaxies} indicate that: i)
Supay and Amun show CAS values which are in agreement with
observations of spiral galaxies \citep{fossati+13,conselice14}; ii)
Siris and Abu show slightly higher values of A, indicating that these
systems have been recently perturbed by mergers, while their C and S
parameters are consistent with late-type disc galaxies.

Overall, the CAS analysis confirms our previous results. Namely, Supay
and Amun are spiral systems while Siris and Abu present a more
complicated morphology.  However, we do not find any obvious
correlation between the formation time of the parent haloes and galaxy
morphologies.
 
\section{Conclusions}
\label{sec:conclusions}

The ZOMG project aims to study the relation between the assembly
history of DM haloes and galaxy formation.  In the present paper we
have concentrated in analysing how gas accretes onto the haloes, and
how this process affects the evolution of their central galaxies. For
this purpose we have performed zoom-in N-body+baryon simulations of
the formation and evolution of four galaxy-sized haloes with masses $M
\sim 5 \times10^{11}\msun$.  The haloes were selected from a larger
sample presented in paper I based on their mass-accretion history. Two
of them continue growing in size until the present time, while the
other two stopped doing so by $z \approx 1$. We refer to the first
group as accreting haloes (Abu and Amun), and to the second one as
stalled haloes (Siris and Supay).

Our main results can be summarised as follows:

\begin{enumerate}

\item DM is acquired via smooth accretion and aggregation of
  satellites. As substructures penetrate the halo, their orbits are
  affected by dynamical friction and their mass content is eroded by
  tidal disruption. These effects reduce the amount of material that
  reaches the central region.  At distances of 300 kpc for $z<1$,
  infall and outflows balance each other for stalled haloes, while
  accreting haloes show recent important mergers.  For both classes no
  net mass inflow is found in the central regions ($r <20$kpc).

\item Although gas experiences hydrodynamical effects (e.g., shock
  heating, stellar feedback), its accretion mimics the DM behaviour at
  the halo outskirts.  

\item In the central regions the distinction between stalled and
  accreting haloes is still present, but to a lesser degree.  The
  former consistently acquire a factor of two less fresh gas (material
  that enters for the first time into the innermost 20 kpc) than the
  latter.  [This holds true also at larger radii up to 100 kpc].  The
  infall rate of recycled material (i.e., gas that has been accreted
  at a given radius $d$, ejected at $r>d$ and then re-accreted)
  decreases with time for stalled haloes during more than half of
  their lifetime. On the other hand, accreting haloes show a constant
  or slight increase in their inflow rates for the same period of
  time. Gas that has been shock heated cools down and `rains' onto the
  central galaxy on relatively long time scales (comparable to the
  Hubble time). For this reason, gas accretion in the central regions
  does not accurately match what happens in the outskirts of the halo.

\item The central galaxies have stellar masses of the order of $1-2
  \times10^{10} \msun$ at the present time, $\sim 2-3$ times smaller
  than the Milky Way \citep{licquia-newman15}. They sit at the peak of
  the star formation efficiency of DM haloes as inferred using abundance
  matching.

\item All galaxies present a stellar disc.  Characterising them into
  their structural components, using either the
  density-profile-fitting method or kinematic decomposition, does not
  show a clear connection between the assembly time of the halo and
  the shape of its central galaxy.  Supay and Amun are disc dominated
  while Abu and Siris are bulge dominated. The two grand-design spiral
  galaxies are found within the haloes with the largest spin
  parameters (see Table~\ref{tab:haloes}). This shows that there is no
  one-to-one correspondence between morphological type and environment
  (stalled haloes are found immerse in filaments while accreting ones
  at the junctions of two or more of them, as shown in paper I).
  Furthermore, observational signatures, drawn from the CAS
  statistics, do not show any clear difference between the different
  halo types.
  
\item The negative metallicity gradients (see
  Table~\ref{tab:galaxies}) indicate that discs formed inside-out. The
  smaller enrichment at the edges suggests that low SF takes place in
  these regions. Fig.~\ref{fig:circularity-t} shows that the
  central parts were formed first, followed by the disc
  components. Present-day SF mainly takes place in the disc, with the
  exception of the disc in Siris in which SF has ceased since $z\sim
  0.2$, but shows some minor contribution at the centre.

\item The median stellar age of the disc component clearly
  reflects the halo assembly time. This quantity is older in the
  stalled haloes, where the disc started to be assembled by
  $z\approx2$. This process is delayed until $z<1$ in the accreting
  haloes.

\item The vertical profiles also present a clear difference between
  the two types of haloes. The stellar discs of the stalled ones are
  thicker by a factor of two with respect to their accreting
  counterparts. This is the result of the reduced supply of fresh gas,
  secular evolution and mostly the action of incoming satellites that
  puff up the stellar discs.

\end{enumerate}

In summary, our study suggests that the median stellar ages and the
vertical profiles of the discs could be used as a proxy for the halo
assembly time. It would be interesting to explore whether assembly
bias is detected when a galaxy population is split in terms of these
variables, at least for the stellar ages since vertical scale-lenghts
are more challenging to measure.

Our results, although lacking statistical support and being focussed
on a narrow halo mass range, present interesting challenges for
galaxy-formation models in which correlations between halo evolution,
gas supply and quenching star-formation in the central galaxy are
directly assumed. In particular, they warn us against making the naive
association between haloes that have collapsed early (late) and red,
early-type (blue, late-type) galaxies.

\section*{Acknowledgements}

We thank Volker Springel for allowing the use of his code
{\sevensize{PGADGET-3}} and the anonymous referee for their
suggestions which improved the presentation of our results.  We
acknowledge support by the Deutsche Forschungsgemeinschaft through the
SFB 956, `The Conditions and Impact of Star Formation' and the TRR 33
`The Dark Universe'.  MB thanks the Bonn-Cologne Graduate School for
Physics and Astronomy for support.  The authors acknowledge that the
results of this research have been achieved using the PRACE-3IP
project (FP7 RI-312763), using the computing resources (Cartesius) at
SURF/SARA, The Netherlands.



\bibliographystyle{mnras}
\bibliography{biblio}

\begin{thebibliography}{}
\makeatletter
\relax
\def\mn@urlcharsother{\let\do\@makeother \do\$\do\&\do\#\do\^\do\_\do\%\do\~}
\def\mn@doi{\begingroup\mn@urlcharsother \@ifnextchar [ {\mn@doi@}
  {\mn@doi@[]}}
\def\mn@doi@[#1]#2{\def\@tempa{#1}\ifx\@tempa\@empty \href
  {http://dx.doi.org/#2} {doi:#2}\else \href {http://dx.doi.org/#2} {#1}\fi
  \endgroup}
\def\mn@eprint#1#2{\mn@eprint@#1:#2::\@nil}
\def\mn@eprint@arXiv#1{\href {http://arxiv.org/abs/#1} {{\tt arXiv:#1}}}
\def\mn@eprint@dblp#1{\href {http://dblp.uni-trier.de/rec/bibtex/#1.xml}
  {dblp:#1}}
\def\mn@eprint@#1:#2:#3:#4\@nil{\def\@tempa {#1}\def\@tempb {#2}\def\@tempc
  {#3}\ifx \@tempc \@empty \let \@tempc \@tempb \let \@tempb \@tempa \fi \ifx
  \@tempb \@empty \def\@tempb {arXiv}\fi \@ifundefined
  {mn@eprint@\@tempb}{\@tempb:\@tempc}{\expandafter \expandafter \csname
  mn@eprint@\@tempb\endcsname \expandafter{\@tempc}}}

\bibitem[\protect\citeauthoryear{{Abadi}, {Navarro}, {Steinmetz}  \&
  {Eke}}{{Abadi} et~al.}{2003}]{abadi+03}
{Abadi} M.~G.,  {Navarro} J.~F.,  {Steinmetz} M.,   {Eke} V.~R.,  2003, \mn@doi
  [\apj] {10.1086/378316}, \href
  {http://adsabs.harvard.edu/abs/2003ApJ...597...21A} {597, 21}

\bibitem[\protect\citeauthoryear{{Adhikari}, {Dalal}  \&
  {Chamberlain}}{{Adhikari} et~al.}{2014}]{adhikari+14}
{Adhikari} S.,  {Dalal} N.,   {Chamberlain} R.~T.,  2014, \mn@doi [\jcap]
  {10.1088/1475-7516/2014/11/019}, \href
  {http://adsabs.harvard.edu/abs/2014JCAP...11..019A} {11, 019}

\bibitem[\protect\citeauthoryear{{Aumer}, {White}, {Naab}  \&
  {Scannapieco}}{{Aumer} et~al.}{2013}]{aumer+13}
{Aumer} M.,  {White} S.~D.~M.,  {Naab} T.,   {Scannapieco} C.,  2013, \mn@doi
  [\mnras] {10.1093/mnras/stt1230}, \href
  {http://adsabs.harvard.edu/abs/2013MNRAS.434.3142A} {434, 3142}

\bibitem[\protect\citeauthoryear{{Baldry} et~al.,}{{Baldry}
  et~al.}{2012}]{baldry+12}
{Baldry} I.~K.,  et~al., 2012, \mn@doi [\mnras]
  {10.1111/j.1365-2966.2012.20340.x}, \href
  {http://adsabs.harvard.edu/abs/2012MNRAS.421..621B} {421, 621}

\bibitem[\protect\citeauthoryear{{Behroozi}, {Wechsler}  \&
  {Conroy}}{{Behroozi} et~al.}{2013a}]{behroozi+13a}
{Behroozi} P.~S.,  {Wechsler} R.~H.,   {Conroy} C.,  2013a, \mn@doi [\apjl]
  {10.1088/2041-8205/762/2/L31}, \href
  {http://adsabs.harvard.edu/abs/2013ApJ...762L..31B} {762, L31}

\bibitem[\protect\citeauthoryear{{Behroozi}, {Wechsler}  \&
  {Conroy}}{{Behroozi} et~al.}{2013b}]{behroozi+13}
{Behroozi} P.~S.,  {Wechsler} R.~H.,   {Conroy} C.,  2013b, \mn@doi [\apj]
  {10.1088/0004-637X/770/1/57}, \href
  {http://adsabs.harvard.edu/abs/2013ApJ...770...57B} {770, 57}

\bibitem[\protect\citeauthoryear{{Benson}}{{Benson}}{2012}]{benson12}
{Benson} A.~J.,  2012, \mn@doi [\na] {10.1016/j.newast.2011.07.004}, \href
  {http://adsabs.harvard.edu/abs/2012NewA...17..175B} {17, 175}

\bibitem[\protect\citeauthoryear{{Birnboim} \& {Dekel}}{{Birnboim} \&
  {Dekel}}{2003}]{birnboim-dekel03}
{Birnboim} Y.,  {Dekel} A.,  2003, \mn@doi [\mnras]
  {10.1046/j.1365-8711.2003.06955.x}, \href
  {http://adsabs.harvard.edu/abs/2003MNRAS.345..349B} {345, 349}

\bibitem[\protect\citeauthoryear{{Borzyszkowski}, {Ludlow}  \&
  {Porciani}}{{Borzyszkowski} et~al.}{2014}]{miko+14}
{Borzyszkowski} M.,  {Ludlow} A.~D.,   {Porciani} C.,  2014, \mn@doi [\mnras]
  {10.1093/mnras/stu2033}, \href
  {http://adsabs.harvard.edu/abs/2014MNRAS.445.4124B} {445, 4124}

\bibitem[\protect\citeauthoryear{{Borzyszkowski}, {Porciani}, {Romano-Diaz}  \&
  {Garaldi}}{{Borzyszkowski} et~al.}{2016}]{miko+16}
{Borzyszkowski} M.,  {Porciani} C.,  {Romano-Diaz} E.,   {Garaldi} E.,  2016,
  preprint, \href {http://adsabs.harvard.edu/abs/2016arXiv161004231B} {}
  (\mn@eprint {arXiv} {1610.04231})

\bibitem[\protect\citeauthoryear{{Bournaud}, {Elmegreen}  \&
  {Martig}}{{Bournaud} et~al.}{2009}]{bournaud+09}
{Bournaud} F.,  {Elmegreen} B.~G.,   {Martig} M.,  2009, \mn@doi [\apjl]
  {10.1088/0004-637X/707/1/L1}, \href
  {http://adsabs.harvard.edu/abs/2009ApJ...707L...1B} {707, L1}

\bibitem[\protect\citeauthoryear{{Bower}, {Benson}, {Malbon}, {Helly}, {Frenk},
  {Baugh}, {Cole}  \& {Lacey}}{{Bower} et~al.}{2006}]{bower+06}
{Bower} R.~G.,  {Benson} A.~J.,  {Malbon} R.,  {Helly} J.~C.,  {Frenk} C.~S.,
  {Baugh} C.~M.,  {Cole} S.,   {Lacey} C.~G.,  2006, \mn@doi [\mnras]
  {10.1111/j.1365-2966.2006.10519.x}, \href
  {http://adsabs.harvard.edu/abs/2006MNRAS.370..645B} {370, 645}

\bibitem[\protect\citeauthoryear{{Bray} et~al.,}{{Bray} et~al.}{2016}]{bray+16}
{Bray} A.~D.,  et~al., 2016, \mn@doi [\mnras] {10.1093/mnras/stv2316}, \href
  {http://adsabs.harvard.edu/abs/2016MNRAS.455..185B} {455, 185}

\bibitem[\protect\citeauthoryear{{Brook}, {Stinson}, {Gibson}, {Shen},
  {Macci{\`o}}, {Obreja}, {Wadsley}  \& {Quinn}}{{Brook}
  et~al.}{2014}]{brook+14}
{Brook} C.~B.,  {Stinson} G.,  {Gibson} B.~K.,  {Shen} S.,  {Macci{\`o}} A.~V.,
   {Obreja} A.,  {Wadsley} J.,   {Quinn} T.,  2014, \mn@doi [\mnras]
  {10.1093/mnras/stu1406}, \href
  {http://adsabs.harvard.edu/abs/2014MNRAS.443.3809B} {443, 3809}

\bibitem[\protect\citeauthoryear{{Brooks} \& {Christensen}}{{Brooks} \&
  {Christensen}}{2016}]{brooks-christensen16}
{Brooks} A.,  {Christensen} C.,  2016, \mn@doi [Galactic Bulges]
  {10.1007/978-3-319-19378-6_12}, \href
  {http://adsabs.harvard.edu/abs/2016ASSL..418..317B} {418, 317}

\bibitem[\protect\citeauthoryear{{Bullock}, {Dekel}, {Kolatt}, {Kravtsov},
  {Klypin}, {Porciani}  \& {Primack}}{{Bullock} et~al.}{2001}]{bullock01ck}
{Bullock} J.~S.,  {Dekel} A.,  {Kolatt} T.~S.,  {Kravtsov} A.~V.,  {Klypin}
  A.~A.,  {Porciani} C.,   {Primack} J.~R.,  2001, \mn@doi [\apj]
  {10.1086/321477}, 555, 240

\bibitem[\protect\citeauthoryear{{Cacciato}, {van den Bosch}, {More}, {Li},
  {Mo}  \& {Yang}}{{Cacciato} et~al.}{2009}]{cacciato+09}
{Cacciato} M.,  {van den Bosch} F.~C.,  {More} S.,  {Li} R.,  {Mo} H.~J.,
  {Yang} X.,  2009, \mn@doi [\mnras] {10.1111/j.1365-2966.2008.14362.x}, \href
  {http://adsabs.harvard.edu/abs/2009MNRAS.394..929C} {394, 929}

\bibitem[\protect\citeauthoryear{{Cacciato}, {van den Bosch}, {More}, {Mo}  \&
  {Yang}}{{Cacciato} et~al.}{2013}]{cacciato+13}
{Cacciato} M.,  {van den Bosch} F.~C.,  {More} S.,  {Mo} H.,   {Yang} X.,
  2013, \mn@doi [\mnras] {10.1093/mnras/sts525}, \href
  {http://adsabs.harvard.edu/abs/2013MNRAS.430..767C} {430, 767}

\bibitem[\protect\citeauthoryear{{Chaves-Montero}, {Angulo}, {Schaye},
  {Schaller}, {Crain}, {Furlong}  \& {Theuns}}{{Chaves-Montero}
  et~al.}{2016}]{chaves-montero+16}
{Chaves-Montero} J.,  {Angulo} R.~E.,  {Schaye} J.,  {Schaller} M.,  {Crain}
  R.~A.,  {Furlong} M.,   {Theuns} T.,  2016, \mn@doi [\mnras]
  {10.1093/mnras/stw1225}, \href
  {http://adsabs.harvard.edu/abs/2016MNRAS.460.3100C} {460, 3100}

\bibitem[\protect\citeauthoryear{{Ciotti} \& {Bertin}}{{Ciotti} \&
  {Bertin}}{1999}]{ciotti-bertin}
{Ciotti} L.,  {Bertin} G.,  1999, \aap, \href
  {http://adsabs.harvard.edu/abs/1999A%26A...352..447C} {352, 447}

\bibitem[\protect\citeauthoryear{{Cole}, {Lacey}, {Baugh}  \& {Frenk}}{{Cole}
  et~al.}{2000}]{cole+00}
{Cole} S.,  {Lacey} C.~G.,  {Baugh} C.~M.,   {Frenk} C.~S.,  2000, \mn@doi
  [\mnras] {10.1046/j.1365-8711.2000.03879.x}, \href
  {http://adsabs.harvard.edu/abs/2000MNRAS.319..168C} {319, 168}

\bibitem[\protect\citeauthoryear{{Conroy} \& {Wechsler}}{{Conroy} \&
  {Wechsler}}{2009}]{conroy+09}
{Conroy} C.,  {Wechsler} R.~H.,  2009, \mn@doi [\apj]
  {10.1088/0004-637X/696/1/620}, \href
  {http://adsabs.harvard.edu/abs/2009ApJ...696..620C} {696, 620}

\bibitem[\protect\citeauthoryear{{Conselice}}{{Conselice}}{2003}]{conselice03}
{Conselice} C.~J.,  2003, \mn@doi [\apjs] {10.1086/375001}, \href
  {http://adsabs.harvard.edu/abs/2003ApJS..147....1C} {147, 1}

\bibitem[\protect\citeauthoryear{{Conselice}}{{Conselice}}{2014}]{conselice14}
{Conselice} C.~J.,  2014, \mn@doi [\araa]
  {10.1146/annurev-astro-081913-040037}, \href
  {http://adsabs.harvard.edu/abs/2014ARA%26A..52..291C} {52, 291}

\bibitem[\protect\citeauthoryear{{Crain}, {Eke}, {Frenk}, {Jenkins},
  {McCarthy}, {Navarro}  \& {Pearce}}{{Crain} et~al.}{2007}]{crain+07}
{Crain} R.~A.,  {Eke} V.~R.,  {Frenk} C.~S.,  {Jenkins} A.,  {McCarthy} I.~G.,
  {Navarro} J.~F.,   {Pearce} F.~R.,  2007, \mn@doi [\mnras]
  {10.1111/j.1365-2966.2007.11598.x}, \href
  {http://adsabs.harvard.edu/abs/2007MNRAS.377...41C} {377, 41}

\bibitem[\protect\citeauthoryear{{Cresci} et~al.,}{{Cresci}
  et~al.}{2015}]{cresci+15}
{Cresci} G.,  et~al., 2015, \mn@doi [\aap] {10.1051/0004-6361/201526581}, \href
  {http://adsabs.harvard.edu/abs/2015A%26A...582A..63C} {582, A63}

\bibitem[\protect\citeauthoryear{{Dalal}, {White}, {Bond}  \&
  {Shirokov}}{{Dalal} et~al.}{2008}]{dalal+08}
{Dalal} N.,  {White} M.,  {Bond} J.~R.,   {Shirokov} A.,  2008, \mn@doi [\apj]
  {10.1086/591512}, \href {http://adsabs.harvard.edu/abs/2008ApJ...687...12D}
  {687, 12}

\bibitem[\protect\citeauthoryear{{Dekel} \& {Birnboim}}{{Dekel} \&
  {Birnboim}}{2006}]{dekel-birnboim06}
{Dekel} A.,  {Birnboim} Y.,  2006, \mn@doi [\mnras]
  {10.1111/j.1365-2966.2006.10145.x}, \href
  {http://adsabs.harvard.edu/abs/2006MNRAS.368....2D} {368, 2}

\bibitem[\protect\citeauthoryear{{Dekel} et~al.,}{{Dekel}
  et~al.}{2009}]{dekel+09}
{Dekel} A.,  et~al., 2009, \mn@doi [\nat] {10.1038/nature07648}, \href
  {http://adsabs.harvard.edu/abs/2009Natur.457..451D} {457, 451}

\bibitem[\protect\citeauthoryear{{Di Matteo}, {Springel}  \& {Hernquist}}{{Di
  Matteo} et~al.}{2005}]{dimatteo+05}
{Di Matteo} T.,  {Springel} V.,   {Hernquist} L.,  2005, \mn@doi [\nat]
  {10.1038/nature03335}, \href
  {http://adsabs.harvard.edu/abs/2005Natur.433..604D} {433, 604}

\bibitem[\protect\citeauthoryear{{Diemer} \& {Kravtsov}}{{Diemer} \&
  {Kravtsov}}{2014}]{diemer+14}
{Diemer} B.,  {Kravtsov} A.~V.,  2014, \mn@doi [\apj]
  {10.1088/0004-637X/789/1/1}, \href
  {http://adsabs.harvard.edu/abs/2014ApJ...789....1D} {789, 1}

\bibitem[\protect\citeauthoryear{{Diemer}, {More}  \& {Kravtsov}}{{Diemer}
  et~al.}{2013}]{diemer+13}
{Diemer} B.,  {More} S.,   {Kravtsov} A.~V.,  2013, \mn@doi [\apj]
  {10.1088/0004-637X/766/1/25}, \href
  {http://adsabs.harvard.edu/abs/2013ApJ...766...25D} {766, 25}

\bibitem[\protect\citeauthoryear{{Fabian}}{{Fabian}}{2012}]{fabian12}
{Fabian} A.~C.,  2012, \mn@doi [\araa] {10.1146/annurev-astro-081811-125521},
  \href {http://adsabs.harvard.edu/abs/2012ARA%26A..50..455F} {50, 455}

\bibitem[\protect\citeauthoryear{{Fakhouri}, {Ma}  \&
  {Boylan-Kolchin}}{{Fakhouri} et~al.}{2010}]{fakhouri+10}
{Fakhouri} O.,  {Ma} C.-P.,   {Boylan-Kolchin} M.,  2010, \mn@doi [\mnras]
  {10.1111/j.1365-2966.2010.16859.x}, \href
  {http://adsabs.harvard.edu/abs/2010MNRAS.406.2267F} {406, 2267}

\bibitem[\protect\citeauthoryear{{Faucher-Gigu{\`e}re}, {Kere{\v s}}  \&
  {Ma}}{{Faucher-Gigu{\`e}re} et~al.}{2011}]{fauchergiguere+11}
{Faucher-Gigu{\`e}re} C.-A.,  {Kere{\v s}} D.,   {Ma} C.-P.,  2011, \mn@doi
  [\mnras] {10.1111/j.1365-2966.2011.19457.x}, \href
  {http://adsabs.harvard.edu/abs/2011MNRAS.417.2982F} {417, 2982}

\bibitem[\protect\citeauthoryear{{Fossati} et~al.,}{{Fossati}
  et~al.}{2013}]{fossati+13}
{Fossati} M.,  et~al., 2013, \mn@doi [\aap] {10.1051/0004-6361/201220915},
  \href {http://adsabs.harvard.edu/abs/2013A%26A...553A..91F} {553, A91}

\bibitem[\protect\citeauthoryear{{Gallazzi}, {Charlot}, {Brinchmann}, {White}
  \& {Tremonti}}{{Gallazzi} et~al.}{2005}]{gallazzi+05}
{Gallazzi} A.,  {Charlot} S.,  {Brinchmann} J.,  {White} S.~D.~M.,   {Tremonti}
  C.~A.,  2005, \mn@doi [\mnras] {10.1111/j.1365-2966.2005.09321.x}, \href
  {http://adsabs.harvard.edu/abs/2005MNRAS.362...41G} {362, 41}

\bibitem[\protect\citeauthoryear{{Gao}, {Springel}  \& {White}}{{Gao}
  et~al.}{2005}]{gao+05}
{Gao} L.,  {Springel} V.,   {White} S.~D.~M.,  2005, \mn@doi [\mnras]
  {10.1111/j.1745-3933.2005.00084.x}, \href
  {http://adsabs.harvard.edu/abs/2005MNRAS.363L..66G} {363, L66}

\bibitem[\protect\citeauthoryear{{Garaldi}, {Romano-D{\'{\i}}az},
  {Borzyszkowski}  \& {Porciani}}{{Garaldi} et~al.}{2016}]{garaldi+16}
{Garaldi} E.,  {Romano-D{\'{\i}}az} E.,  {Borzyszkowski} M.,   {Porciani} C.,
  2016, astroph

\bibitem[\protect\citeauthoryear{{Gill}, {Knebe}  \& {Gibson}}{{Gill}
  et~al.}{2004}]{Gill+04}
{Gill} S.~P.~D.,  {Knebe} A.,   {Gibson} B.~K.,  2004, \mn@doi [\mnras]
  {10.1111/j.1365-2966.2004.07786.x}, \href
  {http://adsabs.harvard.edu/abs/2004MNRAS.351..399G} {351, 399}

\bibitem[\protect\citeauthoryear{{Guo} et~al.,}{{Guo} et~al.}{2011}]{guo+11}
{Guo} Q.,  et~al., 2011, \mn@doi [\mnras] {10.1111/j.1365-2966.2010.18114.x},
  \href {http://adsabs.harvard.edu/abs/2011MNRAS.413..101G} {413, 101}

\bibitem[\protect\citeauthoryear{{Haardt} \& {Madau}}{{Haardt} \&
  {Madau}}{2001}]{haardt-madau01}
{Haardt} F.,  {Madau} P.,  2001, in {Neumann} D.~M.,  {Tran} J.~T.~V.,  eds,
  Clusters of Galaxies and the High Redshift Universe Observed in X-rays.
  (\mn@eprint {} {astro-ph/0106018})

\bibitem[\protect\citeauthoryear{{Hahn} \& {Abel}}{{Hahn} \&
  {Abel}}{2011}]{music}
{Hahn} O.,  {Abel} T.,  2011, \mn@doi [\mnras]
  {10.1111/j.1365-2966.2011.18820.x}, \href
  {http://adsabs.harvard.edu/abs/2011MNRAS.415.2101H} {415, 2101}

\bibitem[\protect\citeauthoryear{{Hearin} \& {Watson}}{{Hearin} \&
  {Watson}}{2013}]{hearin-watson13}
{Hearin} A.~P.,  {Watson} D.~F.,  2013, \mn@doi [\mnras]
  {10.1093/mnras/stt1374}, \href
  {http://adsabs.harvard.edu/abs/2013MNRAS.435.1313H} {435, 1313}

\bibitem[\protect\citeauthoryear{{Hearin}, {Zentner}, {van den Bosch},
  {Campbell}  \& {Tollerud}}{{Hearin} et~al.}{2016}]{Hearin+16}
{Hearin} A.~P.,  {Zentner} A.~R.,  {van den Bosch} F.~C.,  {Campbell} D.,
  {Tollerud} E.,  2016, \mn@doi [\mnras] {10.1093/mnras/stw840}, \href
  {http://adsabs.harvard.edu/abs/2016MNRAS.460.2552H} {460, 2552}

\bibitem[\protect\citeauthoryear{{Henriques}, {White}, {Thomas}, {Angulo},
  {Guo}, {Lemson}, {Springel}  \& {Overzier}}{{Henriques}
  et~al.}{2015}]{henriques+15}
{Henriques} B.~M.~B.,  {White} S.~D.~M.,  {Thomas} P.~A.,  {Angulo} R.,  {Guo}
  Q.,  {Lemson} G.,  {Springel} V.,   {Overzier} R.,  2015, \mn@doi [\mnras]
  {10.1093/mnras/stv705}, \href
  {http://adsabs.harvard.edu/abs/2015MNRAS.451.2663H} {451, 2663}

\bibitem[\protect\citeauthoryear{{Ishibashi} \& {Fabian}}{{Ishibashi} \&
  {Fabian}}{2012}]{ishibashi+12}
{Ishibashi} W.,  {Fabian} A.~C.,  2012, \mn@doi [\mnras]
  {10.1111/j.1365-2966.2012.22074.x}, \href
  {http://adsabs.harvard.edu/abs/2012MNRAS.427.2998I} {427, 2998}

\bibitem[\protect\citeauthoryear{{Kauffmann}, {Li}, {Zhang}  \&
  {Weinmann}}{{Kauffmann} et~al.}{2013}]{kauffmann+13}
{Kauffmann} G.,  {Li} C.,  {Zhang} W.,   {Weinmann} S.,  2013, \mn@doi [\mnras]
  {10.1093/mnras/stt007}, \href
  {http://adsabs.harvard.edu/abs/2013MNRAS.430.1447K} {430, 1447}

\bibitem[\protect\citeauthoryear{{Kazantzidis}, {Bullock}, {Zentner},
  {Kravtsov}  \& {Moustakas}}{{Kazantzidis} et~al.}{2008}]{kazantzidis+08}
{Kazantzidis} S.,  {Bullock} J.~S.,  {Zentner} A.~R.,  {Kravtsov} A.~V.,
  {Moustakas} L.~A.,  2008, \mn@doi [\apj] {10.1086/591958}, \href
  {http://adsabs.harvard.edu/abs/2008ApJ...688..254K} {688, 254}

\bibitem[\protect\citeauthoryear{{Kennicutt}}{{Kennicutt}}{1998}]{kennicutt98}
{Kennicutt} Jr. R.~C.,  1998, \mn@doi [\apj] {10.1086/305588}, \href
  {http://adsabs.harvard.edu/abs/1998ApJ...498..541K} {498, 541}

\bibitem[\protect\citeauthoryear{{Kere{\v s}}, {Katz}, {Weinberg}  \&
  {Dav{\'e}}}{{Kere{\v s}} et~al.}{2005}]{keres+05}
{Kere{\v s}} D.,  {Katz} N.,  {Weinberg} D.~H.,   {Dav{\'e}} R.,  2005, \mn@doi
  [\mnras] {10.1111/j.1365-2966.2005.09451.x}, \href
  {http://adsabs.harvard.edu/abs/2005MNRAS.363....2K} {363, 2}

\bibitem[\protect\citeauthoryear{{Kere{\v s}}, {Katz}, {Fardal}, {Dav{\'e}}  \&
  {Weinberg}}{{Kere{\v s}} et~al.}{2009}]{keres+09a}
{Kere{\v s}} D.,  {Katz} N.,  {Fardal} M.,  {Dav{\'e}} R.,   {Weinberg} D.~H.,
  2009, \mn@doi [\mnras] {10.1111/j.1365-2966.2009.14541.x}, \href
  {http://adsabs.harvard.edu/abs/2009MNRAS.395..160K} {395, 160}

\bibitem[\protect\citeauthoryear{{Knobel}, {Lilly}, {Woo}  \& {Kova{\v
  c}}}{{Knobel} et~al.}{2015}]{knobel+15}
{Knobel} C.,  {Lilly} S.~J.,  {Woo} J.,   {Kova{\v c}} K.,  2015, \mn@doi
  [\apj] {10.1088/0004-637X/800/1/24}, \href
  {http://adsabs.harvard.edu/abs/2015ApJ...800...24K} {800, 24}

\bibitem[\protect\citeauthoryear{{Knollmann} \& {Knebe}}{{Knollmann} \&
  {Knebe}}{2009}]{ahf}
{Knollmann} S.~R.,  {Knebe} A.,  2009, \mn@doi [\apjs]
  {10.1088/0067-0049/182/2/608}, \href
  {http://adsabs.harvard.edu/abs/2009ApJS..182..608K} {182, 608}

\bibitem[\protect\citeauthoryear{{Kravtsov}, {Berlind}, {Wechsler}, {Klypin},
  {Gottl{\"o}ber}, {Allgood}  \& {Primack}}{{Kravtsov}
  et~al.}{2004}]{kravtsov+04}
{Kravtsov} A.~V.,  {Berlind} A.~A.,  {Wechsler} R.~H.,  {Klypin} A.~A.,
  {Gottl{\"o}ber} S.,  {Allgood} B.,   {Primack} J.~R.,  2004, \mn@doi [\apj]
  {10.1086/420959}, \href {http://adsabs.harvard.edu/abs/2004ApJ...609...35K}
  {609, 35}

\bibitem[\protect\citeauthoryear{{Li}, {Mo}  \& {Gao}}{{Li}
  et~al.}{2008}]{li+08}
{Li} Y.,  {Mo} H.~J.,   {Gao} L.,  2008, \mn@doi [\mnras]
  {10.1111/j.1365-2966.2008.13667.x}, \href
  {http://adsabs.harvard.edu/abs/2008MNRAS.389.1419L} {389, 1419}

\bibitem[\protect\citeauthoryear{{Licquia} \& {Newman}}{{Licquia} \&
  {Newman}}{2015}]{licquia-newman15}
{Licquia} T.~C.,  {Newman} J.~A.,  2015, \mn@doi [\apj]
  {10.1088/0004-637X/806/1/96}, \href
  {http://adsabs.harvard.edu/abs/2015ApJ...806...96L} {806, 96}

\bibitem[\protect\citeauthoryear{{MacArthur}, {Courteau}  \&
  {Holtzman}}{{MacArthur} et~al.}{2003}]{macarthur+03}
{MacArthur} L.~A.,  {Courteau} S.,   {Holtzman} J.~A.,  2003, \mn@doi [\apj]
  {10.1086/344506}, \href {http://adsabs.harvard.edu/abs/2003ApJ...582..689M}
  {582, 689}

\bibitem[\protect\citeauthoryear{{Maller}, {Katz}, {Kere{\v s}}, {Dav{\'e}}  \&
  {Weinberg}}{{Maller} et~al.}{2006}]{maller+06}
{Maller} A.~H.,  {Katz} N.,  {Kere{\v s}} D.,  {Dav{\'e}} R.,   {Weinberg}
  D.~H.,  2006, \mn@doi [\apj] {10.1086/503319}, \href
  {http://adsabs.harvard.edu/abs/2006ApJ...647..763M} {647, 763}

\bibitem[\protect\citeauthoryear{{Mandelbaum}, {Seljak}, {Kauffmann}, {Hirata}
  \& {Brinkmann}}{{Mandelbaum} et~al.}{2006}]{mandelbaum+06}
{Mandelbaum} R.,  {Seljak} U.,  {Kauffmann} G.,  {Hirata} C.~M.,   {Brinkmann}
  J.,  2006, \mn@doi [\mnras] {10.1111/j.1365-2966.2006.10156.x}, \href
  {http://adsabs.harvard.edu/abs/2006MNRAS.368..715M} {368, 715}

\bibitem[\protect\citeauthoryear{{Marinacci}, {Pakmor}  \&
  {Springel}}{{Marinacci} et~al.}{2014}]{marinacci+14}
{Marinacci} F.,  {Pakmor} R.,   {Springel} V.,  2014, \mn@doi [\mnras]
  {10.1093/mnras/stt2003}, \href
  {http://adsabs.harvard.edu/abs/2014MNRAS.437.1750M} {437, 1750}

\bibitem[\protect\citeauthoryear{{McGaugh}}{{McGaugh}}{2012}]{McGaugh12}
{McGaugh} S.~S.,  2012, \mn@doi [\aj] {10.1088/0004-6256/143/2/40}, \href
  {http://adsabs.harvard.edu/abs/2012AJ....143...40M} {143, 40}

\bibitem[\protect\citeauthoryear{{McGaugh}, {Schombert}, {de Blok}  \&
  {Zagursky}}{{McGaugh} et~al.}{2010}]{mcgaugh+10}
{McGaugh} S.~S.,  {Schombert} J.~M.,  {de Blok} W.~J.~G.,   {Zagursky} M.~J.,
  2010, \mn@doi [\apjl] {10.1088/2041-8205/708/1/L14}, \href
  {http://adsabs.harvard.edu/abs/2010ApJ...708L..14M} {708, L14}

\bibitem[\protect\citeauthoryear{{Mendez-Abreu} et~al.,}{{Mendez-Abreu}
  et~al.}{2016}]{mendez-abreu+16}
{Mendez-Abreu} J.,  et~al., 2016, preprint, \href
  {http://adsabs.harvard.edu/abs/2016arXiv161005324M} {} (\mn@eprint {arXiv}
  {1610.05324})

\bibitem[\protect\citeauthoryear{{Metuki}, {Libeskind}, {Hoffman}, {Crain}  \&
  {Theuns}}{{Metuki} et~al.}{2015}]{metuki+15}
{Metuki} O.,  {Libeskind} N.~I.,  {Hoffman} Y.,  {Crain} R.~A.,   {Theuns} T.,
  2015, \mn@doi [\mnras] {10.1093/mnras/stu2166}, \href
  {http://adsabs.harvard.edu/abs/2015MNRAS.446.1458M} {446, 1458}

\bibitem[\protect\citeauthoryear{{Minchev}, {Martig}, {Streich}, {Scannapieco},
  {de Jong}  \& {Steinmetz}}{{Minchev} et~al.}{2015}]{minchev+15}
{Minchev} I.,  {Martig} M.,  {Streich} D.,  {Scannapieco} C.,  {de Jong} R.~S.,
    {Steinmetz} M.,  2015, \mn@doi [\apjl] {10.1088/2041-8205/804/1/L9}, \href
  {http://adsabs.harvard.edu/abs/2015ApJ...804L...9M} {804, L9}

\bibitem[\protect\citeauthoryear{{More}, {van den Bosch}, {Cacciato}, {More},
  {Mo}  \& {Yang}}{{More} et~al.}{2013}]{more+13}
{More} S.,  {van den Bosch} F.~C.,  {Cacciato} M.,  {More} A.,  {Mo} H.,
  {Yang} X.,  2013, \mn@doi [\mnras] {10.1093/mnras/sts697}, \href
  {http://adsabs.harvard.edu/abs/2013MNRAS.430..747M} {430, 747}

\bibitem[\protect\citeauthoryear{{More}, {Diemer}  \& {Kravtsov}}{{More}
  et~al.}{2015}]{more+15}
{More} S.,  {Diemer} B.,   {Kravtsov} A.~V.,  2015, \mn@doi [\apj]
  {10.1088/0004-637X/810/1/36}, \href
  {http://adsabs.harvard.edu/abs/2015ApJ...810...36M} {810, 36}

\bibitem[\protect\citeauthoryear{{Moster}, {Naab}  \& {White}}{{Moster}
  et~al.}{2013}]{moster+13}
{Moster} B.~P.,  {Naab} T.,   {White} S.~D.~M.,  2013, \mn@doi [\mnras]
  {10.1093/mnras/sts261}, \href
  {http://adsabs.harvard.edu/abs/2013MNRAS.428.3121M} {428, 3121}

\bibitem[\protect\citeauthoryear{{Murali}, {Katz}, {Hernquist}, {Weinberg}  \&
  {Dav{\'e}}}{{Murali} et~al.}{2002}]{murali+02}
{Murali} C.,  {Katz} N.,  {Hernquist} L.,  {Weinberg} D.~H.,   {Dav{\'e}} R.,
  2002, \mn@doi [\apj] {10.1086/339876}, \href
  {http://adsabs.harvard.edu/abs/2002ApJ...571....1M} {571, 1}

\bibitem[\protect\citeauthoryear{{Peirani}, {Jung}, {Silk}  \&
  {Pichon}}{{Peirani} et~al.}{2012}]{peirani+12}
{Peirani} S.,  {Jung} I.,  {Silk} J.,   {Pichon} C.,  2012, \mn@doi [\mnras]
  {10.1111/j.1365-2966.2012.22105.x}, \href
  {http://adsabs.harvard.edu/abs/2012MNRAS.427.2625P} {427, 2625}

\bibitem[\protect\citeauthoryear{{Peng} et~al.,}{{Peng} et~al.}{2010}]{peng+10}
{Peng} Y.-j.,  et~al., 2010, \mn@doi [\apj] {10.1088/0004-637X/721/1/193},
  \href {http://adsabs.harvard.edu/abs/2010ApJ...721..193P} {721, 193}

\bibitem[\protect\citeauthoryear{{Petrosian}}{{Petrosian}}{1976}]{petrosian76}
{Petrosian} V.,  1976, \mn@doi [\apjl] {10.1086/182253}, \href
  {http://adsabs.harvard.edu/abs/1976ApJ...209L...1P} {209, L1}

\bibitem[\protect\citeauthoryear{{Planck Collaboration} et~al.,}{{Planck
  Collaboration} et~al.}{2014}]{planck+14}
{Planck Collaboration} et~al., 2014, \mn@doi [\aap]
  {10.1051/0004-6361/201321529}, \href
  {http://adsabs.harvard.edu/abs/2014A%26A...571A...1P} {571, A1}

\bibitem[\protect\citeauthoryear{{Purcell}, {Bullock}  \&
  {Kazantzidis}}{{Purcell} et~al.}{2010}]{purcell+10}
{Purcell} C.~W.,  {Bullock} J.~S.,   {Kazantzidis} S.,  2010, \mn@doi [\mnras]
  {10.1111/j.1365-2966.2010.16429.x}, \href
  {http://adsabs.harvard.edu/abs/2010MNRAS.404.1711P} {404, 1711}

\bibitem[\protect\citeauthoryear{{Qu}, {Di Matteo}, {Lehnert}  \& {van
  Driel}}{{Qu} et~al.}{2011}]{qu+11}
{Qu} Y.,  {Di Matteo} P.,  {Lehnert} M.~D.,   {van Driel} W.,  2011, \mn@doi
  [\aap] {10.1051/0004-6361/201015224}, \href
  {http://adsabs.harvard.edu/abs/2011A%26A...530A..10Q} {530, A10}

\bibitem[\protect\citeauthoryear{{Reddick}, {Wechsler}, {Tinker}  \&
  {Behroozi}}{{Reddick} et~al.}{2013}]{reddick+13}
{Reddick} R.~M.,  {Wechsler} R.~H.,  {Tinker} J.~L.,   {Behroozi} P.~S.,  2013,
  \mn@doi [\apj] {10.1088/0004-637X/771/1/30}, \href
  {http://adsabs.harvard.edu/abs/2013ApJ...771...30R} {771, 30}

\bibitem[\protect\citeauthoryear{{Romano-D{\'{\i}}az}, {Shlosman}, {Heller}  \&
  {Hoffman}}{{Romano-D{\'{\i}}az} et~al.}{2009}]{erd-I}
{Romano-D{\'{\i}}az} E.,  {Shlosman} I.,  {Heller} C.,   {Hoffman} Y.,  2009,
  \mn@doi [\apj] {10.1088/0004-637X/702/2/1250}, \href
  {http://adsabs.harvard.edu/abs/2009ApJ...702.1250R} {702, 1250}

\bibitem[\protect\citeauthoryear{{Romano-D{\'{\i}}az}, {Shlosman}, {Heller}  \&
  {Hoffman}}{{Romano-D{\'{\i}}az} et~al.}{2010}]{erd-II}
{Romano-D{\'{\i}}az} E.,  {Shlosman} I.,  {Heller} C.,   {Hoffman} Y.,  2010,
  \mn@doi [\apj] {10.1088/0004-637X/716/2/1095}, \href
  {http://adsabs.harvard.edu/abs/2010ApJ...716.1095R} {716, 1095}

\bibitem[\protect\citeauthoryear{{Romano-D{\'{\i}}az}, {Shlosman}, {Choi}  \&
  {Sadoun}}{{Romano-D{\'{\i}}az} et~al.}{2014}]{erd+14}
{Romano-D{\'{\i}}az} E.,  {Shlosman} I.,  {Choi} J.-H.,   {Sadoun} R.,  2014,
  \mn@doi [\apjl] {10.1088/2041-8205/790/2/L32}, \href
  {http://adsabs.harvard.edu/abs/2014ApJ...790L..32R} {790, L32}

\bibitem[\protect\citeauthoryear{{Sales}, {Navarro}, {Theuns}, {Schaye},
  {White}, {Frenk}, {Crain}  \& {Dalla Vecchia}}{{Sales}
  et~al.}{2012}]{sales+12}
{Sales} L.~V.,  {Navarro} J.~F.,  {Theuns} T.,  {Schaye} J.,  {White} S.~D.~M.,
   {Frenk} C.~S.,  {Crain} R.~A.,   {Dalla Vecchia} C.,  2012, \mn@doi [\mnras]
  {10.1111/j.1365-2966.2012.20975.x}, \href
  {http://adsabs.harvard.edu/abs/2012MNRAS.423.1544S} {423, 1544}

\bibitem[\protect\citeauthoryear{{Scannapieco}, {Tissera}, {White}  \&
  {Springel}}{{Scannapieco} et~al.}{2008}]{scannapieco+08}
{Scannapieco} C.,  {Tissera} P.~B.,  {White} S.~D.~M.,   {Springel} V.,  2008,
  \mn@doi [\mnras] {10.1111/j.1365-2966.2008.13678.x}, \href
  {http://adsabs.harvard.edu/abs/2008MNRAS.389.1137S} {389, 1137}

\bibitem[\protect\citeauthoryear{{Scannapieco} et~al.,}{{Scannapieco}
  et~al.}{2012}]{Scannapieco+12}
{Scannapieco} C.,  et~al., 2012, \mn@doi [\mnras]
  {10.1111/j.1365-2966.2012.20993.x}, \href
  {http://adsabs.harvard.edu/abs/2012MNRAS.423.1726S} {423, 1726}

\bibitem[\protect\citeauthoryear{{S{\'e}rsic}}{{S{\'e}rsic}}{1963}]{sersic63}
{S{\'e}rsic} J.~L.,  1963, Boletin de la Asociacion Argentina de Astronomia La
  Plata Argentina, \href {http://adsabs.harvard.edu/abs/1963BAAA....6...41S}
  {6, 41}

\bibitem[\protect\citeauthoryear{{Sheth} \& {Tormen}}{{Sheth} \&
  {Tormen}}{2004}]{sheth-tormen04}
{Sheth} R.~K.,  {Tormen} G.,  2004, \mn@doi [\mnras]
  {10.1111/j.1365-2966.2004.07733.x}, \href
  {http://adsabs.harvard.edu/abs/2004MNRAS.350.1385S} {350, 1385}

\bibitem[\protect\citeauthoryear{{Sijacki}, {Springel}, {Di Matteo}  \&
  {Hernquist}}{{Sijacki} et~al.}{2007}]{sijacki+07}
{Sijacki} D.,  {Springel} V.,  {Di Matteo} T.,   {Hernquist} L.,  2007, \mn@doi
  [\mnras] {10.1111/j.1365-2966.2007.12153.x}, \href
  {http://adsabs.harvard.edu/abs/2007MNRAS.380..877S} {380, 877}

\bibitem[\protect\citeauthoryear{{Somerville} \& {Primack}}{{Somerville} \&
  {Primack}}{1999}]{somerville-primack99}
{Somerville} R.~S.,  {Primack} J.~R.,  1999, \mn@doi [\mnras]
  {10.1046/j.1365-8711.1999.03032.x}, \href
  {http://adsabs.harvard.edu/abs/1999MNRAS.310.1087S} {310, 1087}

\bibitem[\protect\citeauthoryear{{Somerville}, {Hopkins}, {Cox}, {Robertson}
  \& {Hernquist}}{{Somerville} et~al.}{2008}]{somerville+08}
{Somerville} R.~S.,  {Hopkins} P.~F.,  {Cox} T.~J.,  {Robertson} B.~E.,
  {Hernquist} L.,  2008, \mn@doi [\mnras] {10.1111/j.1365-2966.2008.13805.x},
  \href {http://adsabs.harvard.edu/abs/2008MNRAS.391..481S} {391, 481}

\bibitem[\protect\citeauthoryear{{Springel}}{{Springel}}{2005}]{gadget}
{Springel} V.,  2005, \mn@doi [\mnras] {10.1111/j.1365-2966.2005.09655.x},
  \href {http://adsabs.harvard.edu/abs/2005MNRAS.364.1105S} {364, 1105}

\bibitem[\protect\citeauthoryear{{Springel} \& {Hernquist}}{{Springel} \&
  {Hernquist}}{2002}]{springel-hernquist+02}
{Springel} V.,  {Hernquist} L.,  2002, \mn@doi [\mnras]
  {10.1046/j.1365-8711.2002.05445.x}, \href
  {http://adsabs.harvard.edu/abs/2002MNRAS.333..649S} {333, 649}

\bibitem[\protect\citeauthoryear{{Springel} \& {Hernquist}}{{Springel} \&
  {Hernquist}}{2003}]{springel-hernquist+03}
{Springel} V.,  {Hernquist} L.,  2003, \mn@doi [\mnras]
  {10.1046/j.1365-8711.2003.06206.x}, \href
  {http://adsabs.harvard.edu/abs/2003MNRAS.339..289S} {339, 289}

\bibitem[\protect\citeauthoryear{{Tinker}, {Weinberg}, {Zheng}  \&
  {Zehavi}}{{Tinker} et~al.}{2005}]{tinker+05}
{Tinker} J.~L.,  {Weinberg} D.~H.,  {Zheng} Z.,   {Zehavi} I.,  2005, \mn@doi
  [\apj] {10.1086/432084}, \href
  {http://adsabs.harvard.edu/abs/2005ApJ...631...41T} {631, 41}

\bibitem[\protect\citeauthoryear{{Tremonti} et~al.,}{{Tremonti}
  et~al.}{2004}]{tremonti+04}
{Tremonti} C.~A.,  et~al., 2004, \mn@doi [\apj] {10.1086/423264}, \href
  {http://adsabs.harvard.edu/abs/2004ApJ...613..898T} {613, 898}

\bibitem[\protect\citeauthoryear{{Vale} \& {Ostriker}}{{Vale} \&
  {Ostriker}}{2004}]{vale-ostriker04}
{Vale} A.,  {Ostriker} J.~P.,  2004, \mn@doi [\mnras]
  {10.1111/j.1365-2966.2004.08059.x}, \href
  {http://adsabs.harvard.edu/abs/2004MNRAS.353..189V} {353, 189}

\bibitem[\protect\citeauthoryear{{Villaescusa-Navarro}, {Marulli}, {Viel},
  {Branchini}, {Castorina}, {Sefusatti}  \& {Saito}}{{Villaescusa-Navarro}
  et~al.}{2014}]{villaescusa-navarro+14}
{Villaescusa-Navarro} F.,  {Marulli} F.,  {Viel} M.,  {Branchini} E.,
  {Castorina} E.,  {Sefusatti} E.,   {Saito} S.,  2014, \mn@doi [\jcap]
  {10.1088/1475-7516/2014/03/011}, \href
  {http://adsabs.harvard.edu/abs/2014JCAP...03..011V} {3, 011}

\bibitem[\protect\citeauthoryear{{Villalobos} \& {Helmi}}{{Villalobos} \&
  {Helmi}}{2008}]{villalobos08}
{Villalobos} {\'A}.,  {Helmi} A.,  2008, \mn@doi [\mnras]
  {10.1111/j.1365-2966.2008.13979.x}, \href
  {http://adsabs.harvard.edu/abs/2008MNRAS.391.1806V} {391, 1806}

\bibitem[\protect\citeauthoryear{{Vogelsberger}, {White}, {Mohayaee}  \&
  {Springel}}{{Vogelsberger} et~al.}{2009}]{vogelsberger+09}
{Vogelsberger} M.,  {White} S.~D.~M.,  {Mohayaee} R.,   {Springel} V.,  2009,
  \mn@doi [\mnras] {10.1111/j.1365-2966.2009.15615.x}, \href
  {http://adsabs.harvard.edu/abs/2009MNRAS.400.2174V} {400, 2174}

\bibitem[\protect\citeauthoryear{{Wechsler}, {Zentner}, {Bullock}, {Kravtsov}
  \& {Allgood}}{{Wechsler} et~al.}{2006}]{wechsler+06}
{Wechsler} R.~H.,  {Zentner} A.~R.,  {Bullock} J.~S.,  {Kravtsov} A.~V.,
  {Allgood} B.,  2006, \mn@doi [\apj] {10.1086/507120}, \href
  {http://adsabs.harvard.edu/abs/2006ApJ...652...71W} {652, 71}

\bibitem[\protect\citeauthoryear{{Weinmann}, {van den Bosch}, {Yang}  \&
  {Mo}}{{Weinmann} et~al.}{2006}]{weinmann+06}
{Weinmann} S.~M.,  {van den Bosch} F.~C.,  {Yang} X.,   {Mo} H.~J.,  2006,
  \mn@doi [\mnras] {10.1111/j.1365-2966.2005.09865.x}, \href
  {http://adsabs.harvard.edu/abs/2006MNRAS.366....2W} {366, 2}

\bibitem[\protect\citeauthoryear{{White} \& {Frenk}}{{White} \&
  {Frenk}}{1991}]{white-frenk91}
{White} S.~D.~M.,  {Frenk} C.~S.,  1991, \mn@doi [\apj] {10.1086/170483}, \href
  {http://adsabs.harvard.edu/abs/1991ApJ...379...52W} {379, 52}

\bibitem[\protect\citeauthoryear{{White} \& {Rees}}{{White} \&
  {Rees}}{1978}]{white-rees78}
{White} S.~D.~M.,  {Rees} M.~J.,  1978, \mn@doi [\mnras]
  {10.1093/mnras/183.3.341}, \href
  {http://adsabs.harvard.edu/abs/1978MNRAS.183..341W} {183, 341}

\bibitem[\protect\citeauthoryear{{Yepes}, {Kates}, {Khokhlov}  \&
  {Klypin}}{{Yepes} et~al.}{1997}]{yepes+97}
{Yepes} G.,  {Kates} R.,  {Khokhlov} A.,   {Klypin} A.,  1997, \mn@doi [\mnras]
  {10.1093/mnras/284.1.235}, \href
  {http://adsabs.harvard.edu/abs/1997MNRAS.284..235Y} {284, 235}

\bibitem[\protect\citeauthoryear{{Zemp}}{{Zemp}}{2014}]{zemp14}
{Zemp} M.,  2014, \mn@doi [\apj] {10.1088/0004-637X/792/2/124}, \href
  {http://adsabs.harvard.edu/abs/2014ApJ...792..124Z} {792, 124}

\bibitem[\protect\citeauthoryear{{Zolotov}, {Willman}, {Brooks}, {Governato},
  {Hogg}, {Shen}  \& {Wadsley}}{{Zolotov} et~al.}{2010}]{zolotov+10}
{Zolotov} A.,  {Willman} B.,  {Brooks} A.~M.,  {Governato} F.,  {Hogg} D.~W.,
  {Shen} S.,   {Wadsley} J.,  2010, \mn@doi [\apj]
  {10.1088/0004-637X/721/1/738}, \href
  {http://adsabs.harvard.edu/abs/2010ApJ...721..738Z} {721, 738}

\bibitem[\protect\citeauthoryear{{van de Voort}, {Schaye}, {Booth}, {Haas}  \&
  {Dalla Vecchia}}{{van de Voort} et~al.}{2011}]{vandevoort+11}
{van de Voort} F.,  {Schaye} J.,  {Booth} C.~M.,  {Haas} M.~R.,   {Dalla
  Vecchia} C.,  2011, \mn@doi [\mnras] {10.1111/j.1365-2966.2011.18565.x},
  \href {http://adsabs.harvard.edu/abs/2011MNRAS.414.2458V} {414, 2458}

\bibitem[\protect\citeauthoryear{{van der Kruit}}{{van der
  Kruit}}{1988}]{vanderkruit88}
{van der Kruit} P.~C.,  1988, \aap, \href
  {http://adsabs.harvard.edu/abs/1988A%26A...192..117V} {192, 117}

\makeatother
\end{thebibliography}


\bsp	
\label{lastpage}
\end{document}